\documentclass[cernpreprint, orcidlogo, atlasdraft=false, texlive=2023, british, texmf, orcidlogo]{atlasdoc}
 
\usepackage[subfigure]{atlaspackage}
 
\usepackage{atlasbiblatex}
\usepackage{atlasphysics}

\graphicspath{{logos/}{figures/}}
 
\usepackage{ANA-EXOT-2021-15-PAPER-defs}
\usepackage{tablefootnote}
\usepackage{multirow}
\usepackage{pifont}

\AtlasTitle{Search for pair production of third-generation leptoquarks decaying into a bottom quark
and a $\tau$-lepton with the ATLAS detector}

\AtlasJournalRef{Eur. Phys. J. C 83 (2023) 1075}
\AtlasDOI{10.1140/epjc/s10052-023-12104-7}
 
\AtlasAbstract{
A search for pair-produced scalar or vector leptoquarks decaying into a $b$-quark and a
$\tau$-lepton is presented using the full LHC Run~2 (2015--2018) data sample of 139\,fb$^{-1}$
collected with the ATLAS detector in proton--proton collisions at a centre-of-mass energy of
$\sqrt{s} =13$\,TeV.  Events in which at least one $\tau$-lepton decays hadronically are considered,
and multivariate discriminants are used to extract the signals.  No significant deviations from the
Standard Model expectation are observed and 95\% confidence-level upper limits on the production
cross-section are derived as a function of leptoquark mass and branching ratio \BR\ into a
$\tau$-lepton and $b$-quark.  For scalar leptoquarks, masses below 1460\,GeV are excluded assuming
$\BR=100$\%, while for vector leptoquarks the corresponding limit is 1650\,GeV (1910\,GeV) in the
minimal-coupling (Yang--Mills) scenario.  }
 
\author{The ATLAS Collaboration}
 
\journal{EPJC}
 
\AtlasRefCode{EXOT-2021-15}
 
\PreprintIdNumber{CERN-EP-2022-267}

\hypersetup{pdftitle={ATLAS document},pdfauthor={The ATLAS Collaboration}}
 
\begin{document}
 
\maketitle

\section{Introduction}
\label{sec:intro}

Many extensions of the Standard Model (SM) of particle physics predict particles known as
leptoquarks (LQs)~\cite{Theory1,Theory2,Theory3,Theory4,Theory5,Theory6,Theory7}.  These particles
provide a connection between the lepton and quark sectors, which are similar in structure in the
SM\@. LQs can be scalar (spin-0) or vector (spin-1) bosons, and they carry colour and a fractional
electric charge. They also have non-zero lepton and baryon numbers, and decay into quark--lepton pairs.
They can mediate neutral and charge currents, and therefore can  potentially provide an explanation for hints
of violations of lepton universality observed in flavour experiments~\cite{Flavor1,Flavor2,Flavor3,Flavor4,Flavor5,Flavor6,Flavor7}.

This analysis searches for the pair-production of LQs that couple strongly to the third generation
of quarks and leptons in proton--proton ($pp$) collisions at the LHC.  Within the
Buchm\"uller–R\"uckl–Wyler (BRW) model~\cite{Theory8}, which is the benchmark for scalar LQs in this
analysis, it is assumed that these LQs can only interact within the same family via a Yukawa
interaction. This interaction is described by two parameters, a model parameter $\beta$ and a
coupling parameter $\lambda$. In the BRW model, the pair-production cross-section is independent of
$\lambda$. This analysis also considers pair-production of vector LQs~\cite{Blumlein:1996qp,vLQ}
corresponding to the $U_1$ state in the BRW classification~\cite{Theory8}. The scenarios considered
in this model differ by a dimensionless coupling constant $k$, which is zero for the
minimal-coupling scenario and one for the Yang--Mills scenario.  For both scalar and vector LQs, the
parameter $\beta$ controls the coupling to charged leptons.  For third generation LQs, results are
generally given in terms of the mass of the LQ (\mlq) and its branching ratio (\BR), rather than
$\beta$ as is the case for first- and second-generation LQs.  This is because $\beta$ is not equal
to \BR\ for third-generation LQs due to the sizable top-quark mass.

ATLAS and CMS have published searches for LQs coupling to the first, second and third
generations~\cite{EXOT-2019-19,EXOT-2017-30,EXOT-2019-15,SUSY-2018-12,SUSY-2018-34,CMS-EXO-19-015,CMS-EXO-17-009,CMS-EXO-17-016,CMS-EXO-17-003,PreviousSUSY}.
Each generation of LQs is split into up-type and down-type
LQs with different electric charges.  For instance, for the third generation
they are split into up-type LQs ($\text{LQ}^{\text{u}}_{3}$), which decay into $b\tau$ or $t\nu$, and down-type LQs
($\text{LQ}^{\text{d}}_{3}$), which decay into $t\tau$ or $b\nu$. Both types of LQs are currently
excluded for masses below 1150~\GeV\ for the BRW model, for all values of \BR.

\begin{figure}[tbh]
\centering
\includegraphics[width=0.45\textwidth]{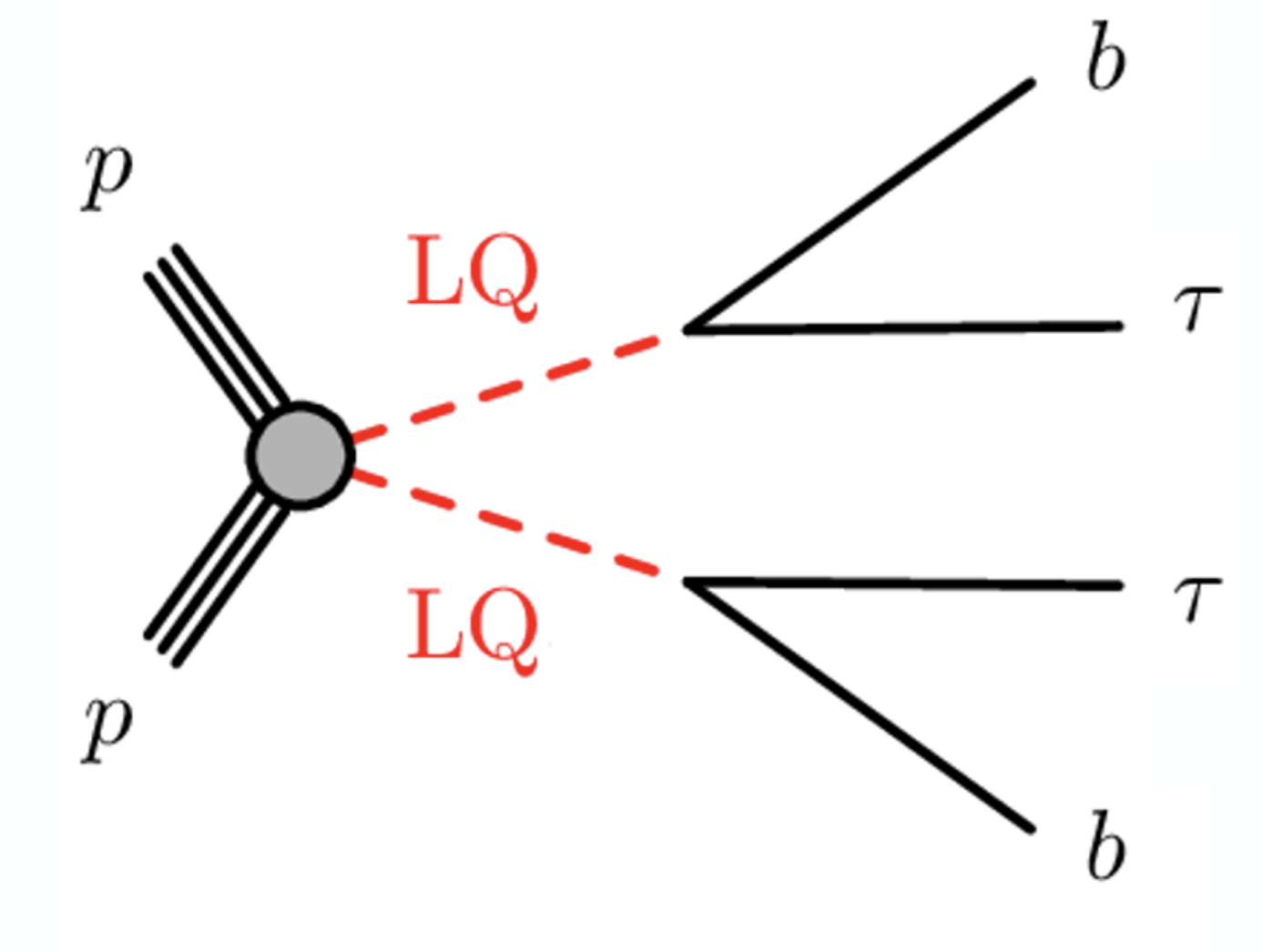}
\caption{Pair production of a leptoquark (LQ) and its subsequent decay into a $b$-quark and a $\tau$-lepton.}
\label{fig:FD}
\end{figure}

This paper updates the ATLAS search for an up-type LQ pair decaying into $b\tau$~\cite{EXOT-2017-30},
shown in Figure~\ref{fig:FD}, using the full Run~2 data sample and an updated analysis strategy,
prioritising high LQ masses that are not yet excluded in the benchmark models considered. Analysis
improvements include updated analysis-optimisation and background-estimation methods, as well as
updates to several object identification algorithms. The analysis signature is two jets, at least
one of which must be identified as containing a $b$-hadron, and two $\tau$-leptons.  For the
$\tau$-leptons, the cases considered are where both decay hadronically or where one
$\tau$-lepton decays into a light lepton (electron or muon, $\ell$) and neutrinos and the other decays
hadronically. The mass range considered for the LQ is from 300\,\GeV\ to 2000\,\GeV. The extraction of
the signals is performed through a simultaneous likelihood fit to multivariate discriminants. For
the results, both scalar and vector LQs are considered, with the limits on vector LQs interpreted in
the context of two scenarios, the Yang--Mills scenario and the minimal-coupling
scenario~\cite{LQscenarios}.
 
The paper is structured as follows. After a brief description of the ATLAS detector, the data
sample, simulated backgrounds and simulated signals are described. This is followed by a description
of the event reconstruction, the object selection, the event
selections for the signal regions, and the multivariate discriminants that are used in the
final fit. The next sections include a description of the data-driven background estimation methods, the
systematic uncertainties, and finally the statistical methods and results.


\section{ATLAS detector}
\label{sec:detector}

\newcommand{\AtlasCoordFootnote}{ATLAS uses a right-handed coordinate system with its origin at
the nominal interaction point (IP) in the centre of the detector and the \(z\)-axis along the beam
pipe.  The \(x\)-axis points from the IP to the centre of the LHC ring, and the \(y\)-axis points
upwards.  Cylindrical coordinates \((r,\phi)\) are used in the transverse plane, \(\phi\) being
the azimuthal angle around the \(z\)-axis.  The pseudorapidity is defined in terms of the polar
angle \(\theta\) as \(\eta = -\ln \tan(\theta/2)\).  Angular distance is measured in units of
\(\Delta R \equiv \sqrt{(\Delta\eta)^{2} + (\Delta\phi)^{2}}\).}

The ATLAS detector~\cite{PERF-2007-01} at the LHC is a multipurpose particle detector with a
forward--backward symmetric cylindrical geometry and a near \(4\pi\) coverage in solid
angle.\footnote{\AtlasCoordFootnote}
The inner tracking detector consists of pixel and microstrip silicon detectors covering the
pseudorapidity region $|\eta| < 2.5$, surrounded by a transition radiation tracker to enhance
electron identification in the range of $|\eta| < 2.0$.  An additional innermost pixel layer, the insertable
B-layer~\cite{ATLAS-TDR-19,PIX-2018-001}, was added before Run 2 of the LHC.
The inner detector (ID) is surrounded by a thin superconducting solenoid providing a \SI{2}{\tesla}
axial magnetic field, and by a fine-granularity lead/liquid-argon (LAr) electromagnetic (EM)
calorimeter covering $|\eta| < 3.2$.  Hadronic calorimetry is provided by a steel/scintillator-tile calorimeter in the central pseudorapidity range ($|\eta| < 1.7$).  The endcap and forward regions are
instrumented with LAr calorimeters for both the EM and hadronic energy measurements up to $|\eta| =
4.9$.
The muon spectrometer (MS) surrounds the calorimeters and is based on three large superconducting
air-core toroidal magnets with eight coils each.  Three layers of high-precision tracking chambers
provide coverage in the range of $|\eta| < 2.7$, while dedicated fast chambers allow triggering in the
region $|\eta| < 2.4$.
A two-level trigger system~\cite{TRIG-2016-01}, consisting of a hardware-based first-level trigger
followed by a software-based high-level trigger (HLT), is used to select events.
An extensive software suite~\cite{ATL-SOFT-PUB-2021-001} is used in data simulation, in the reconstruction and analysis
of real and simulated data, in detector operations, and in the trigger and data acquisition systems
of the experiment.


\section{Data and simulation samples}
\label{sec:samples}

The data used in this search correspond to an integrated luminosity of \lumi{} of $pp$ collision
data collected by the ATLAS detector between 2015 and 2018 at a centre-of-mass energy
$\sqrt{s}=13$\,\TeV.  The uncertainty in the combined 2015--2018 integrated luminosity is
1.7\%~\cite{ATLAS-CONF-2019-021}, obtained using the LUCID-2 detector~\cite{LUCID2} for the primary
luminosity measurements.  The presence of additional interactions in the same or neighbouring bunch
crossing, referred to as \pileup, is characterised by the average number of such interactions,
$\avg{\mu}$, which was $33.7$ for the combined data sample.  Only events recorded under stable beam
conditions and for which all relevant detector subsystems were known to be in a good operating
condition are used.
 
Dedicated Monte Carlo (MC) simulated samples are used to model SM processes and estimate the
expected signal yields.  All samples were passed through the full ATLAS detector
simulation~\cite{SOFT-2010-01} based on \GEANT~\cite{Agostinelli:2002hh}, except for the signal
samples that use a parameterised fast simulation of the calorimeter
response~\cite{ATL-PHYS-PUB-2010-013} and \GEANT for the other detector systems.  The simulated
events were reconstructed with the same algorithms as used for data, and contain a realistic
modelling of pile-up interactions.  The  pile-up profiles match those of each data sample
between 2015 and 2018, and are obtained by overlaying minimum-bias events simulated using the soft QCD
processes of Pythia 8.186~\cite{Sjostrand:2007gs} with the NNPDF2.3 leading-order
(LO)~\cite{Ball:2012cx} set of parton distribution functions (PDFs) and the
A3~\cite{ATL-PHYS-PUB-2016-017} set of tuned parameters (tune).  The MC samples are corrected to
account for the differences between simulation and data in terms of the \pileup, the energy and
momentum scales, and the reconstruction and identification efficiencies of physics objects.

Simulated events with pair-produced up-type ($Q = +\frac{2}{3}$) scalar LQs were generated at
next-to-leading order (NLO) in QCD with \MGNLO{}~v2.6.0~\cite{Alwall:2014hca}, using the LQ model of
Ref.~\cite{sLQ}, in which fixed-order NLO QCD calculations ~\cite{Kramer:2004df,Kramer:1997hh} are
interfaced to \PYTHIA{}~8.230~\cite{Sjostrand:2014zea} for the parton shower (PS) and hadronisation.
Parton luminosities were provided by the five-flavour scheme NNPDF3.0 NLO~\cite{Ball:2014uwa} PDF set
with $\alpha_s = 0.118$ and the underlying event (UE) was modelled with the A14
tune~\cite{ATL-PHYS-PUB-2014-021,ATLAS:2012uec}.
The coupling parameter $\lambda$ was set to 0.3, resulting in a relative LQ width of approximately
0.2\% and ensuring the LQs decay promptly.  In all cases, $\beta = 0.5$ such that the couplings to
charged leptons and neutrinos were equal and the decay products were interfaced
to \MADSPIN~\cite{Artoisenet:2012st} to preserve spin correlations.  Different values for \BR\
were then obtained by reweighting the simulated events according to the generator information about
their decay following the procedure in Ref.~\cite{EXOT-2017-30}.
Signal cross-sections were obtained from the calculation of the pair production of scalar coloured
particles, such as the hypothesised supersymmetric partner of the top quark, as these particles have the same
production modes and their pair-production cross-section depends only on their mass.  These
processes were computed at approximate next-to-next-to-leading order (NNLO) in QCD with resummation
of next-to-next-to-leading-logarithmic (NNLL) soft gluon terms
~\cite{Beenakker:2016lwe,Beenakker:1997ut,Beenakker:2010nq,Beenakker:2016gmf}. The cross-sections do
not include contributions from $t$-channel lepton exchange, which are neglected in Ref.~\cite{sLQ} and may lead to
corrections at the percent level~\cite{Borschensky:2020hot}.  The nominal cross-section and its
uncertainty were derived using the PDF4LHC15\_mc PDF set, following the recommendations of
Ref.~\cite{Butterworth:2015oua}.  For LQ masses between 300\,\GeV\ and 2000\,\GeV, the cross-sections
range from 10\,pb to 0.01\,fb.

Simulated events with pair-produced up-type vector LQs were generated at LO in QCD
with \MGNLO{}~v2.6.0, using the LQ model of Ref.~\cite{vLQ} and the NNPDF3.0 NLO PDF set with
$\alpha_s = 0.118$.
Decays of the LQs were performed with \MADSPIN{}, while PS and hadronisation were simulated
using \PYTHIA{}~8.244 with the A14 tune.
The full model includes two additional vector states that are necessary to obtain a realistic extension
of the SM, a colour singlet $Z^{\prime}$ and a colour octet $G^{\prime}$.  However, these are not
present in the \textsc{MadGraph} model and hence do not contribute to the Feynman diagrams considered for
pair production of vector leptoquarks.
The samples were produced with a coupling strength $g_U = 3.0$, where $g_U$ represents the overall
coupling between the LQ and the fermion, motivated by a suppression of the production cross-section
for the additional mediators in the ultraviolet completion of the model, which might otherwise be in
tension with existing LHC limits.  This choice of coupling results in a relative LQ width of around 10\%.  In
all cases, $\beta = 0.5$ and the same reweighting as in the scalar LQ case is then used to probe
different \BR\ values.
As mentioned, the model introduces two different coupling scenarios,
the minimal-coupling scenario
and the Yang--Mills scenario.
In the latter case the LQ is a massive gauge boson and has additional couplings to the SM gauge
bosons, resulting in enhanced cross-sections.  \BR\ is assumed to be unaffected by these
couplings since the corresponding decays are either forbidden or heavily suppressed.  Since no
higher-order cross-sections are available for this model, the LO \MGNLO{} cross-sections were used
and vary between 94\,pb (340\,pb) and 0.05\,fb (0.61\,fb) for LQ masses between 300\,\GeV\ and
2000\,\GeV\ in the minimal-coupling (Yang--Mills) case.  Above 500\,\GeV, kinematic differences
between the two scenarios are negligible.
 
Scalar (vector) LQ samples were produced with LQ masses between 300\,\GeV\ to 2000\,\GeV, with a mass
interval of 50\,\GeV\ in the range of 800--1600\,\GeV\ (1400--1600\,\GeV) and 100\,\GeV\ otherwise.

\begin{table}[tb]
\caption{The list of generators used for the simulation of the SM background processes.  Information is
given on the matrix element (ME) generator (including the perturbative QCD order), the PDF set,
the parton shower (PS) and the underlying event (UE).  The perturbative order (in QCD unless
otherwise specified) of the cross-section used to normalise the different samples is also
presented.  $(\S)$ The $t\bar{t}-Wt$ interference was handled using the diagram removal
scheme.  $({\dagger})$ The cross-sections from \textsc{Sherpa} at NLO were used to normalise
the $WW,WZ,ZZ$ and $t\bar{t}W/Z$ events. $({\ddagger})$ The $qq\to ZH$ process was normalised to
the NNLO (QCD) + NLO( EW) cross-section for the $pp \to ZH$
process~\cite{Ciccolini:2003jy,Brein:2003wg,Ferrera:2011bk,Brein:2011vx,Ferrera:2014lca,Campbell:2016jau},
after subtracting the $gg\to ZH$ contribution. \protect\label{tab:samples}}
\begin{center}{\small
\resizebox{\textwidth}{!}{\begin{tabular}{llllllll}
\toprule
\hsp Process & ME generator & ME QCD & ME PDF & PS and & UE & Cross-section \hspace{2.5cm}\\
& & order & & hadronisation & tune & order\\
\midrule
\multicolumn{7}{l}{\BF{Top-quark}} \\
\hsp $t\bar{t}^{(\S)}$  & \textsc{Powheg-Box~v2}~\cite{Alioli:2010xd} & NLO & NNPDF3.0NLO& \textsc{Pythia~8.230}  & A14 & NNLO+NNLL~\cite{Czakon:2011xx} \\
\hsp $t$-channel & \textsc{Powheg-Box~v2} & NLO & NNPDF3.0NLO & \textsc{Pythia~8.230} & A14 & NLO~\cite{stop_sch_Xsec} \\
\hsp $s$-channel & \textsc{Powheg-Box~v2} & NLO & NNPDF3.0NLO & \textsc{Pythia~8.230} & A14 & NLO~\cite{stop_tch_Xsec} \\
\hsp $Wt^{(\S)}$        & \textsc{Powheg-Box~v2} & NLO & NNPDF3.0NLO & \textsc{Pythia~8.230} & A14 & NLO~\cite{Kidonakis:2010ux} \\
\midrule
\multicolumn{7}{l}{\BF{Top-quark + $W/Z$}} \\
\hsp $t\bar{t}Z$ & \textsc{Sherpa~2.2.1}~\cite{Bothmann:2019yzt,Gleisberg:2008fv,Cascioli:2011va}  & NLO & NNPDF3.0NNLO &  \textsc{Sherpa~2.2.1} & Default & NLO$^{(\dagger)}$  \\
\hsp $t\bar{t}W$ & \textsc{Sherpa~2.2.8}  & NLO & NNPDF3.0NNLO &  \textsc{Sherpa~2.2.8} & Default & NLO$^{(\dagger)}$  \\
\midrule
\multicolumn{7}{l}{\BF{Vector boson + jets}} \\
\hsp $W/Z+$jets & \textsc{Sherpa~2.2.1} & NLO ($\leq 2$ jets) & NNPDF3.0NNLO &  \textsc{Sherpa~2.2.1} & Default & NNLO~\cite{Catani:2009sm}  \\
&                       & LO  (3,4 jets)     &              &                        &                 \\
\midrule
\multicolumn{7}{l}{\BF{Diboson}} \\
\hsp $WW,WZ,ZZ$ & \textsc{Sherpa~2.2.1} & NLO ($\leq 1$ jet)  & NNPDF3.0NNLO &  \textsc{Sherpa~2.2.1} & Default & NLO$^{(\dagger)}$  \\
&                       & LO (2,3 jets)       &              &                        &         &                     \\
\midrule
\multicolumn{7}{l}{\BF{Higgs boson}} \\
\hsp ggF         & \textsc{Powheg-Box~v2} & NNLO & NNPDF3.0NLO &  \textsc{Pythia~8.212} &
AZNLO~\cite{AZNLOtune} & N3LO(QCD)+NLO(EW)~\cite{deFlorian:2016spz,Anastasiou:2016cez,Anastasiou:2015ema,Dulat:2018rbf,Actis:2008ug}\\
\hsp VBF         & \textsc{Powheg-Box~v2} & NLO  & NNPDF3.0NLO &  \textsc{Pythia~8.212} & AZNLO &
NNLO(QCD)+NLO(EW)~\cite{deFlorian:2016spz,Ciccolini:2007jr,Ciccolini:2007ec,Bolzoni:2010xr}\\
\hsp $qq\to WH$  & \textsc{Powheg-Box~v2} & NLO  & NNPDF3.0NLO &  \textsc{Pythia~8.212} & AZNLO &
NNLO(QCD)+NLO(EW)~\cite{Ciccolini:2003jy,Brein:2003wg,Ferrera:2011bk,Brein:2011vx,Ferrera:2013yga,Campbell:2016jau}  \\
\hsp $qq\to ZH$  & \textsc{Powheg-Box~v2} & NLO  & NNPDF3.0NLO &  \textsc{Pythia~8.212} & AZNLO & NNLO(QCD)+NLO(EW)$^{(\ddagger)}$  \\
\hsp $gg\to ZH$  & \textsc{Powheg-Box~v2} & NLO  & NNPDF3.0NLO &  \textsc{Pythia~8.212} & AZNLO &
NLO+NLL~\cite{Altenkamp:2012sx,Hespel:2015zea,ggzhnll,Harlander:2013mla,Brein:2012ne} \\
\hsp $t\bar{t}H$ & \textsc{Powheg-Box~v2} & NLO  & NNPDF3.0NLO &  \textsc{Pythia~8.230} & A14   & NLO~\cite{deFlorian:2016spz}  \\
\bottomrule
\end{tabular}}}
\end{center}
\end{table}


Background samples were simulated using different MC event generators depending on the process. All
background processes are normalised to the most accurate available theoretical calculation of their
respective cross-sections.  The most relevant event generators, the accuracy of theoretical cross-sections, the UE
parameter tunes, and the PDF sets used in simulating the SM background processes are summarised in Table~\ref{tab:samples}. For all samples, except those generated
using \SHERPA, the \EVTGEN~v1.2.0~\cite{Lange:2001uf} program was used to simulate the properties of
the $b$- and $c$-hadron decays.


\section{Event reconstruction and object definitions}
\label{sec:objects}

 
The LQ signature of interest in this search gives rise to a set of reconstructed objects that
consist primarily of $\tau$-leptons, which may decay into light leptons or hadronically, and jets
from the hadronisation of quarks, specifically $b$-quarks.  In addition, neutrinos produced in the
decay of $\tau$-leptons and the semileptonic decay of $b$-hadrons contribute to the missing
transverse momentum ${\pmb p_{\text{T}}^{\text{miss}}}$ of the event.  To be considered for
analysis, events are required to have at least one $pp$ interaction vertex, reconstructed from two
or more charged-particle tracks with transverse momentum $\pt > 500$\,\MeV; the one with the highest
summed $\pt^2$ of associated tracks is selected as the primary vertex.
 
Electron candidates are reconstructed by matching ID tracks to energy clusters in the EM
calorimeter.  They must satisfy $\pt > 7$\,\GeV\ and lie in the range of $|\eta| < 2.47$, excluding
the transition region between the barrel and endcap detectors ($1.37 < | \eta| < 1.52$).  Electrons
are further identified using a likelihood-based method, based on the track quality, the profile of
the shower measured in the EM calorimeter and the consistency between the track and the energy
cluster~\cite{EGAM-2018-01}.  Two identification criteria are used to select electrons in this
analysis: `veto electrons'\footnote{\label{ftnote}`Veto' leptons are used to reject events with
additional leptons as discussed in \Sect{\ref{sec:sr}}} are required to satisfy the `loose'
identification working point, while `signal electrons' are required to satisfy the more stringent
`tight' working point.

Muon candidates are reconstructed from tracks in the MS, matched with compatible tracks in the ID
where coverage allows; in regions where the MS is only partially instrumented ($|\eta| < 0.1$) an
energy deposit in the calorimeter compatible with a minimum-ionising particle is combined with a
compatible ID track instead.  They must satisfy $\pt > 7$\,\GeV\ and lie in the range of $|\eta| <
2.7$.  Muons are further identified based on the number of hits in the various ID subdetectors and
MS stations, the compatibility between the measurements in the two detectors and the properties of
the resulting track fit.  Two identification criteria~\cite{MUON-2018-03} are used to select muons:
`veto muons'\fnref{ftnote} must satisfy a `loose' identification requirement, while the `signal muons' are
required to satisfy the `medium' (`high-\pt') working point if the $\pt$ is less than (greater than)
800\,\GeV.  The more stringent high-\pt requirements remove around $20\%$ of muons but improve the
$\pt$ resolution by $\approx 30\%$ above 1.5\,\TeV, significantly suppressing potential
backgrounds~\cite{PERF-2015-10}.

To suppress misidentified light leptons or those arising from hadron decays, all light-lepton candidates
must satisfy an isolation criterion that limits the presence of tracks (calorimeter deposits) in a
$\pt$-dependent (fixed) radius cone.  The resulting efficiency is above 99\% for both electrons and
muons in the signal regions.  Finally, signal leptons must satisfy stricter requirements on their
$\pt$ depending on the data-taking period, as detailed in \Sect{\ref{sec:selection}}.
 
Jets are reconstructed from topological energy clusters and charged-particle tracks, resulting from
a particle-flow algorithm~\cite{PERF-2015-09}, using the anti-$k_t$ algorithm with a radius
parameter of $R=0.4$~\cite{Cacciari:2008gp,Fastjet}.  They are required to satisfy $\pt > 20$\,\GeV\
and lie in the range of $|\eta| < 2.5$.  To suppress jets from \pileup, jets with $\pt <
60$\,\GeV\ and $|\eta| < 2.4$ are required to originate from the primary vertex using a multivariate
`jet vertex tagger'~\cite{PERF-2014-03}.
A multivariate algorithm based on a deep neural network, known as the `DL1r
tagger'~\cite{FTAG-2018-01,ATL-PHYS-PUB-2017-013,FTAG-2019-07}, is used to identify jets containing
$b$-hadrons ($b$-jets) based on the jet kinematics, the impact parameters of tracks associated with
the jet and the reconstruction of displaced vertices.  This analysis uses a working point with a
77\% efficiency for true $b$-jets, as measured in simulated $t\bar{t}$ events, and corresponding
rejection factors\footnote{The rejection factor is defined as the reciprocal of the efficiency to
mistag a jet not containing $B$-hadrons as a $b$-jet.} for light-flavour jets, charm jets and $\tau$-leptons of 170, 5 and 21,
respectively~\cite{ATLAS-CONF-2018-006,FTAG-2020-08}.

Hadronically decaying $\tau$-lepton candidates are seeded by jets, which are required to have one or
three associated tracks (referred to hereafter as `one-prong' or `three-prong' candidates,
respectively) with a total charge of $\pm 1$~\cite{PERF-2013-06}.  The
visible decay products ($\tauhadvis$) must satisfy $\pt > 20$\,\GeV\ and lie in the range of $|\eta| <
2.47$, excluding the transition region defined above.  True $\tauhadvis$ candidates are
discriminated from quark- and gluon-initiated jets via a recurrent neural network (RNN) using
calorimeter- and tracking-based variables as input and trained separately on one- and three-prong
candidates~\cite{ATL-PHYS-PUB-2019-033}.  The `loose' working point used has an
efficiency of approximately $85\%$ and $75\%$ for one- and three-prong $\tauhadvis$ respectively.  A
further boosted decision tree (BDT) is used to reject one-prong $\tauhadvis$ candidates originating
from electrons with an efficiency of about $95\%$~\cite{ATLAS-CONF-2017-029}.  For the estimation of
the background from jets misidentified as $\tauhadvis$ (described in \Sect{\ref{sec:background}}),
anti-$\tauhadvis$ candidates are defined in the same way as above but are required to fail to satisfy the nominal
loose RNN working point requirements and instead satisfy a looser requirement that has an efficiency of $99\%$ for
selecting true $\tauhadvis$ candidates.
 
The ${\pmb p_{\text{T}}^{\text{miss}}}$ (with magnitude $\met$) is computed from the negative
vectorial sum of the selected and calibrated objects described above, along with an extra
track-based `soft term' to account for the energy of particles originating from the primary vertex
but not associated to any of the reconstructed objects~\cite{PERF-2016-07,ATLAS-CONF-2018-023}.
 
To resolve ambiguities whereby the same detector signature may be reconstructed as more than one
physics object, a sequential overlap-removal procedure is applied.  First, electron candidates are
discarded if they share a track with a more energetic electron or a muon identified in the MS; if
the muon is identified in the calorimeter it is removed instead.  Any $\tauhadvis$ candidate within
$\Delta R = 0.2$ of an electron or a muon (which must be reconstructed in the MS if the $\tauhadvis$
$\pt$ is above 50\,\GeV) is then rejected.  Jets are discarded if they lie within $\Delta R = 0.2$ of
an electron or have fewer than three associated tracks and lie within the same distance of a muon.
Electron or muon ($\tauhadvis$) candidates within $\Delta R = 0.4$ ($\Delta R = 0.2$) of any
remaining jet are then removed.  Finally, ambiguities between anti-$\tauhadvis$ candidates and jets
within $\Delta R = 0.2$ are resolved in favour of the jet if it is $b$-tagged or the
anti-$\tauhadvis$ otherwise.


\section{Event selection}
\label{sec:selection}

The event selection targets a signature consisting of a pair of $\tau$-leptons and a pair of
$b$-quarks.  It splits the events into two orthogonal signal categories based on the $\tau$-lepton decay
mode: the \lephad\ channel, which selects events with a light lepton, an oppositely charged
\tauhadvis\ and one or two $b$-jets, and the \hadhad\ channel, which selects events with two
opposite-charge \tauhadvis\ and one or two $b$-jets.  Multivariate techniques are used to search for a
LQ-pair signal in the two signal regions (SRs).

\subsection{Signal regions}
\label{sec:sr}
 
Candidate events were recorded using a combination of single-light-lepton~\cite{TRIG-2018-05,TRIG-2018-01}
and single-\tauhadvis\ triggers~\cite{ATLAS-CONF-2017-061}.  The single-lepton trigger used in
the \lephad\ channel required a reconstructed light lepton at the HLT, with a minimum $E_{\text{T}}$
threshold ranging from 24 to 26\,\GeV\ for electrons and a minimum \pt\ threshold ranging from 20 to
25\,\GeV\ for the muons, depending on the data-taking period.  Offline leptons are required to be
geometrically matched to the corresponding trigger object and have a \pt threshold 1--2\,\GeV\ above
the HLT threshold so that the trigger was fully efficient.  The single-\tauhadvis\
triggers used in the \hadhad\ channel required a reconstructed HLT \tauhadvis\ with a
period-dependent minimum \pt\ threshold ranging between 80\,\GeV\ and 160\,\GeV.  The
corresponding \pt-threshold for the offline \tauhadvis, which is again required to be geometrically
matched to the trigger object, ranges between 100\,\GeV\ and 180\,\GeV, while the
non-trigger-matched \tauhadvis\ is required to have $\pt > 20$~\GeV.
 
Following the trigger selection, the \lephad\ category requires exactly one `signal' light lepton and an
oppositely charged \tauhadvis, while the \hadhad\ category requires exactly two opposite-charge
\tauhadvis\ and no `veto' light leptons.  Both categories require at least two jets, one or two of
which must be $b$-tagged, with $\pt > 45~(20)$\,\GeV\ for the leading (sub-leading) jet.
 
The invariant mass of the two $\tau$-lepton decay products is an important variable with which to reject the $Z$+jets
background.  It is calculated using the missing mass calculator (MMC)~\cite{Elagin:2010aw}, with the
light lepton and the \tauhadvis\ (two \tauhadvis) and the ${\pmb p_{\text{T}}^{\text{miss}}}$ as
input in the \lephad\ (\hadhad) category, and it is required to satisfy $\mmc \notin 40 - 150$\,\GeV.  Two
further selections are applied to target the characteristic LQ signature while reducing the large
multi-jet background.  The scalar sum of the transverse momenta (\st),
calculated taking into account the light lepton or \tauhadvis, two leading jets and the \met,
is a powerful discriminator.  It is required to satisfy $\st > 600$\,\GeV, while the \met itself is
required to exceed 100\,\GeV.
 
\begin{figure}
\centering
\subfigure[]{\includegraphics[width=0.45\textwidth]{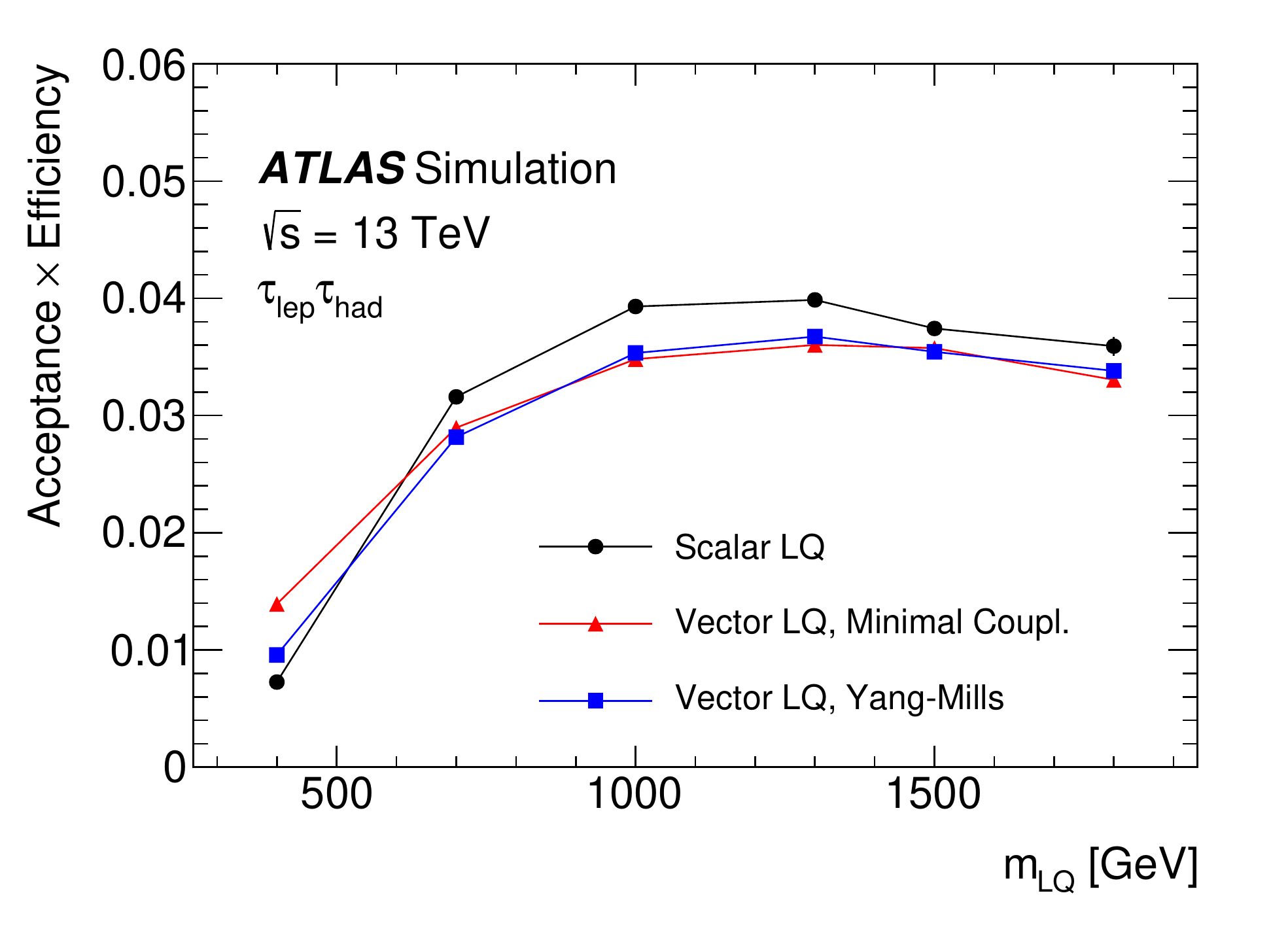}}
\subfigure[]{\includegraphics[width=0.45\textwidth]{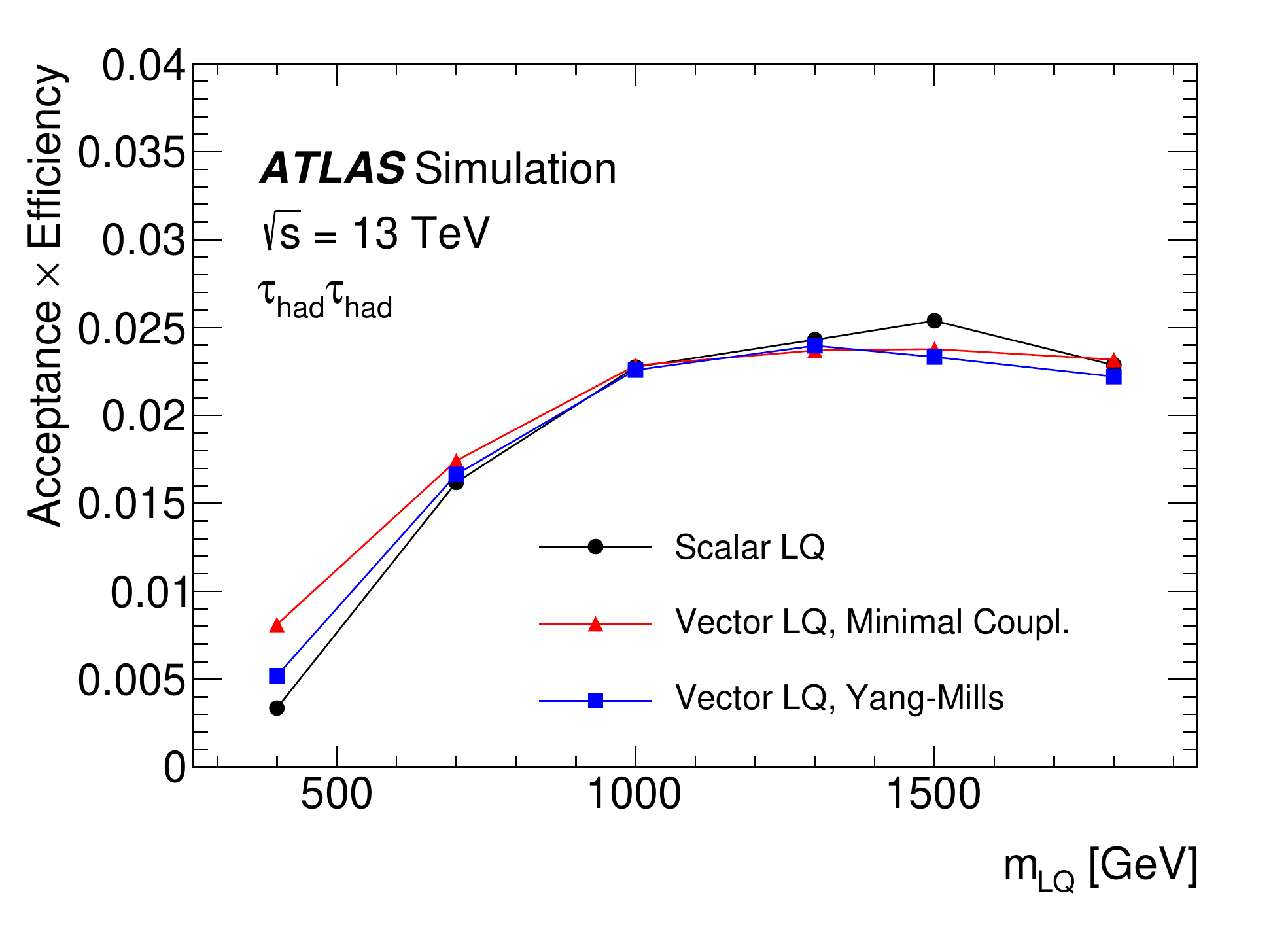}}
 
\caption{The expected acceptance times efficiency (including object identification and reconstruction,
triggering, and event selection) for the scalar and vector LQs, with both the minimal-coupling
and the Yang--Mills scenarios, at $\beta = 0.5$ as a function of \mlq\ in the (a) \lephad\ and (b) \hadhad\
channels.  The values include the leptonic and hadronic branching ratios of the tau
lepton.  The error bars, which are in general smaller than the markers, indicate the statistical
uncertainty.}
\label{fig:acc}
\end{figure}

The full event selection is summarised in Table~\ref{tab:cuts} and the resulting acceptance times
efficiency is shown in Figure~\ref{fig:acc} as a function of \mlq.  Since
the analysis prioritises high mass LQs that have not yet been excluded in the benchmark models under
consideration, it is not optimal for low LQ masses.

 
\begin{table}[tbh]
\begin{center}
\caption{Summary of the event selections for the \lephad\ and \hadhad\ categories. Where two
objects are required, the thresholds on the sub-leading object are given
in parenthesis.  Where the selection depends on data-taking period, the
different possible threshold values are separated by commas.\protect\label{tab:cuts}}
\begin{tabular}{lccc}
\toprule
 
& \lephad\ channel & \hadhad\ channel \\
 
\midrule
 
\multirow{3}{*}{$e/\mu$ selection}
& $=1$ `signal' $e$ or $\mu$ & No `veto' $e$ or $\mu$ \\
& $\pt^e > 25,27$\,\GeV    &   \\
& $\pt^{\mu} > 21,27$\,\GeV &  & \\ [+0.1cm]
 
\hline
\multirow{2}{*}{$\tauhadvis$ selection}
& $=1$ \tauhadvis & $=2$ \tauhadvis \\
& $\pt^{\tau} > 100$\,\GeV & $\pt^{\tau} > 100, 140, 180~(20)$\,\GeV \\ [+0.1cm]
 
\hline
\multirow{3}{*}{Jet selection}
& \multicolumn{2}{c}{$\ge 2$ jets} \\
& \multicolumn{2}{c}{$\pt^{\text{jet}} > 45~(20)$\,\GeV} \\
& \multicolumn{2}{c}{1 or 2  $b$-jets} \\ [+0.1cm]
 
\hline
\multirow{4}{*}{Additional selection}
& \multicolumn{2}{c}{Opposite charge $e,\mu,\tauhad$ and $\tauhad$} \\
& \multicolumn{2}{c}{$\mmc \notin 40 - 150$\,\GeV} \\
& \multicolumn{2}{c}{$\met > 100$\,\GeV}  \\
& \multicolumn{2}{c}{$\st > 600$\,\GeV}  \\
 
\bottomrule
 
\end{tabular}
\end{center}
\end{table}


\subsection{Multivariate signal extraction}
 
Following the event selection, the LQ signal is extracted using a multivariate discriminant.
To obtain near-optimal sensitivity and continuity over the full range of LQ masses
considered, a parameterised neural network (PNN)~\cite{Baldi:2016fzo}, parameterised in terms of the
generated LQ mass, is chosen.  The PNN consists of three hidden layers, each with 32 nodes,
implemented in Keras~\cite{keras} with the Tensorflow ~\cite{tensorflow} backend.

The PNN inputs consist of a combination of multiplicity, kinematic and angular quantities that
discriminate between the signal and the dominant background.  In the case of the $b\tau$ invariant mass, the most
likely combination of the $\tau$-lepton and a $b$-jet\footnote{In the case of only one $b$-jet, the
highest-\pt\ non-$b$-jet is taken as the second jet.} is chosen based on a mass-pairing strategy
that minimises the mass difference between the two resulting LQ candidates.  The variables, which
are similar for both the \lephad\ and \hadhad\ categories, are summarised in Table~\ref{tab:mva} and
defined as follows:
\begin{itemize}
\item $\tauhadvis$ $\pt^{0}$ is the transverse momentum of the highest-\pt\ \tauhadvis;
\item $\st$ is the scalar sum of the transverse momenta defined above;
\item $N_{b-\text{jets}}$ is the number of $b$-jets;
 
\item $m(\tau, \text{jet})_{0,1}$ are the larger (0) and smaller (1) of the two LQ masses obtained
via the mass-pairing strategy (\hadhad\ channel only);
\item $m(\ell, \text{jet})$ and $m(\tauhad, \text{jet})$ are the mass of the light-lepton or
\tauhadvis, respectively, combined with its mass-paired $b$-jet (\lephad\ channel only);
\item $\Delta R(\ell, \text{jet})$ ($\Delta R(\tauhad, \text{jet})$)
is the $\Delta R$ between the light lepton (leading \tauhadvis)
and the mass-paired jet in the \lephad\ (\hadhad) category;
\item $\Delta \phi(\ell, \met)$ is the azimuthal opening angle between the lepton and the \met
(\lephad\ category only);
\item $\met$ $\phi$ centrality quantifies the transverse direction of the ${\pmb
p_{\text{T}}^{\text{miss}}}$ relative to the light lepton and \tauhadvis\ (two \tauhadvis) in
the \lephad\ (\hadhad) category and is defined in Ref.~\cite{HIGG-2016-16}.
\end{itemize}
 
\begin{figure}
\centering
\subfigure[]{\includegraphics[width=.49\textwidth]{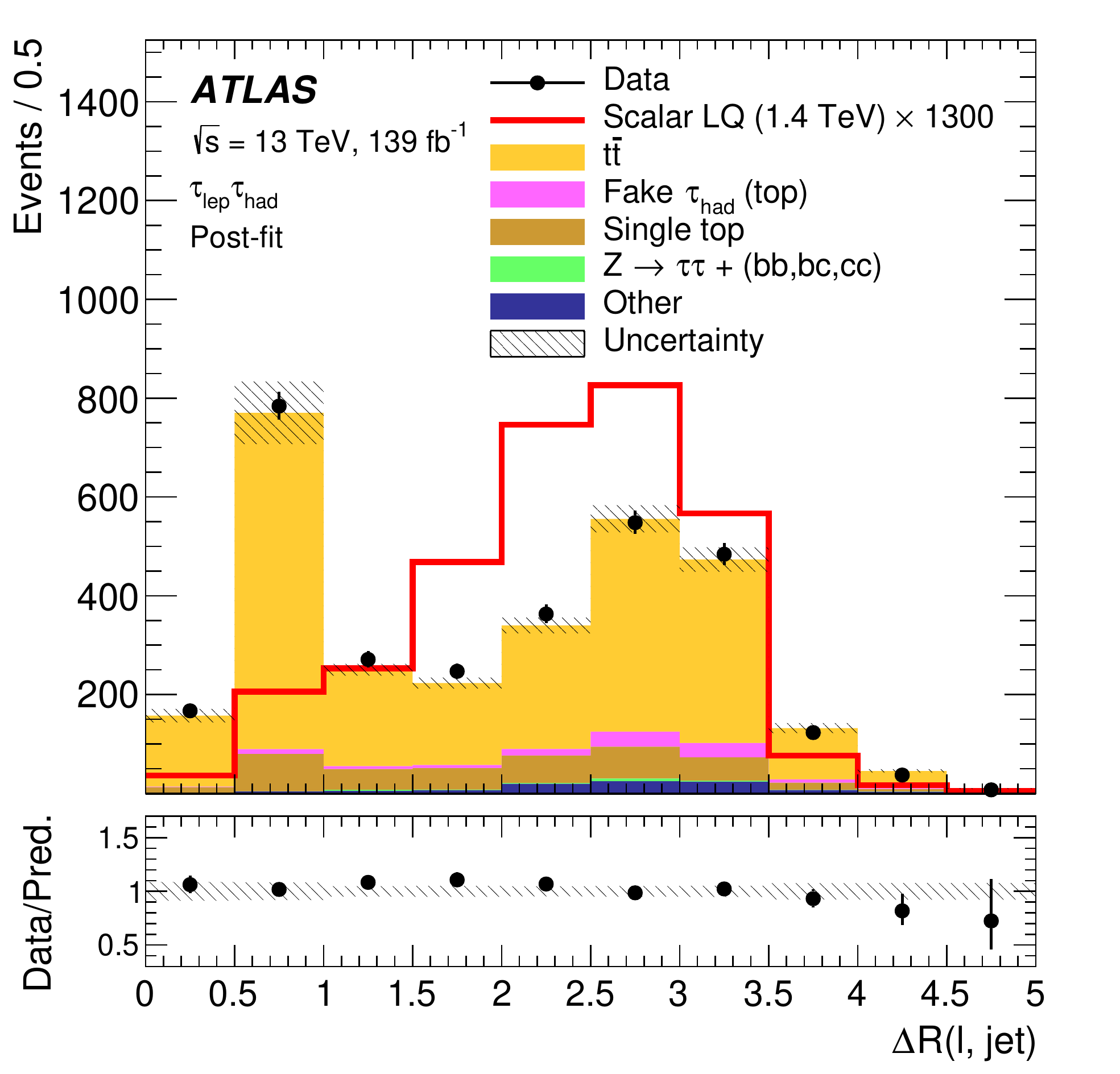}}
\subfigure[]{\includegraphics[width=.49\textwidth]{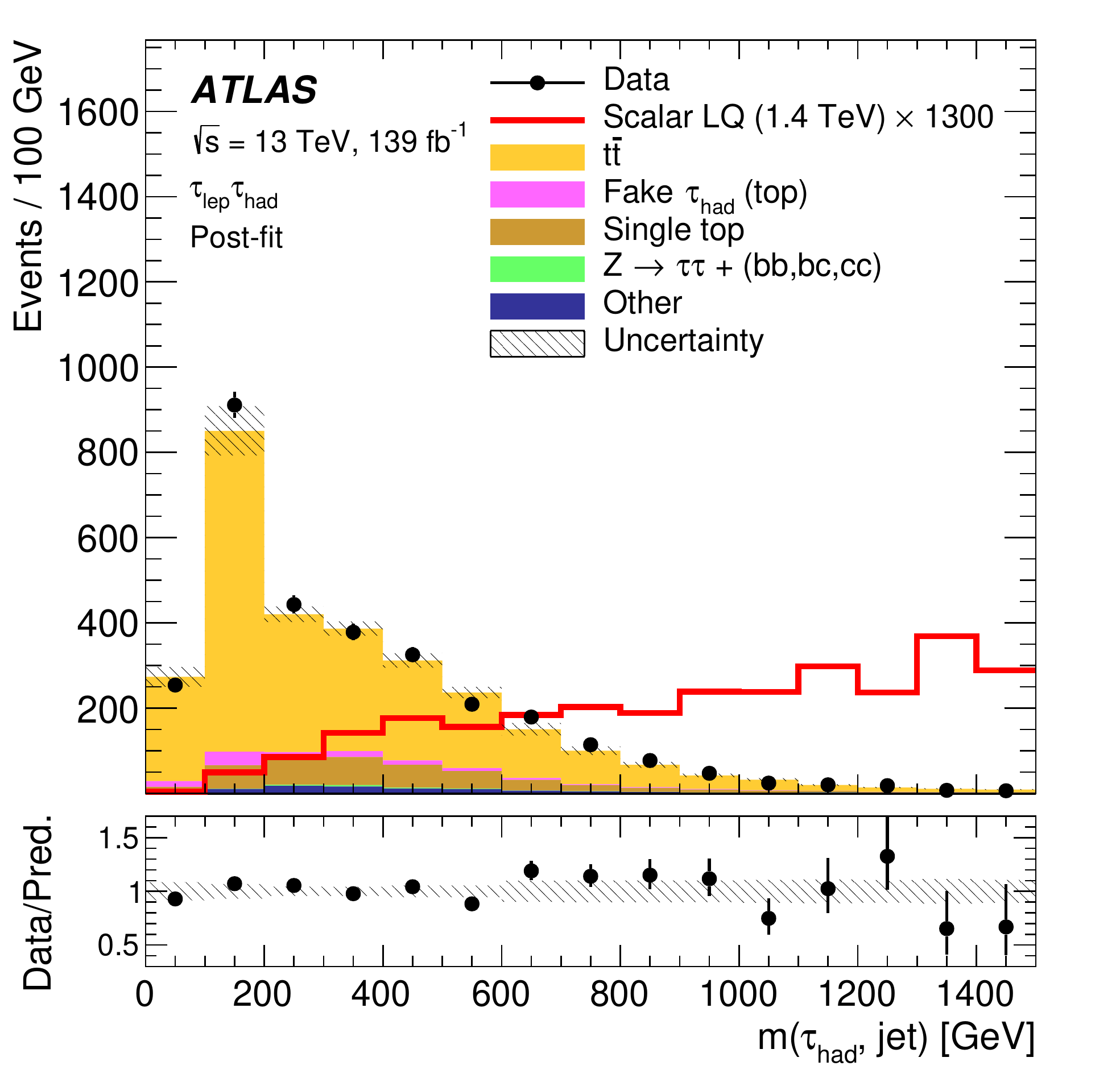}}
\subfigure[]{\includegraphics[width=.49\textwidth]{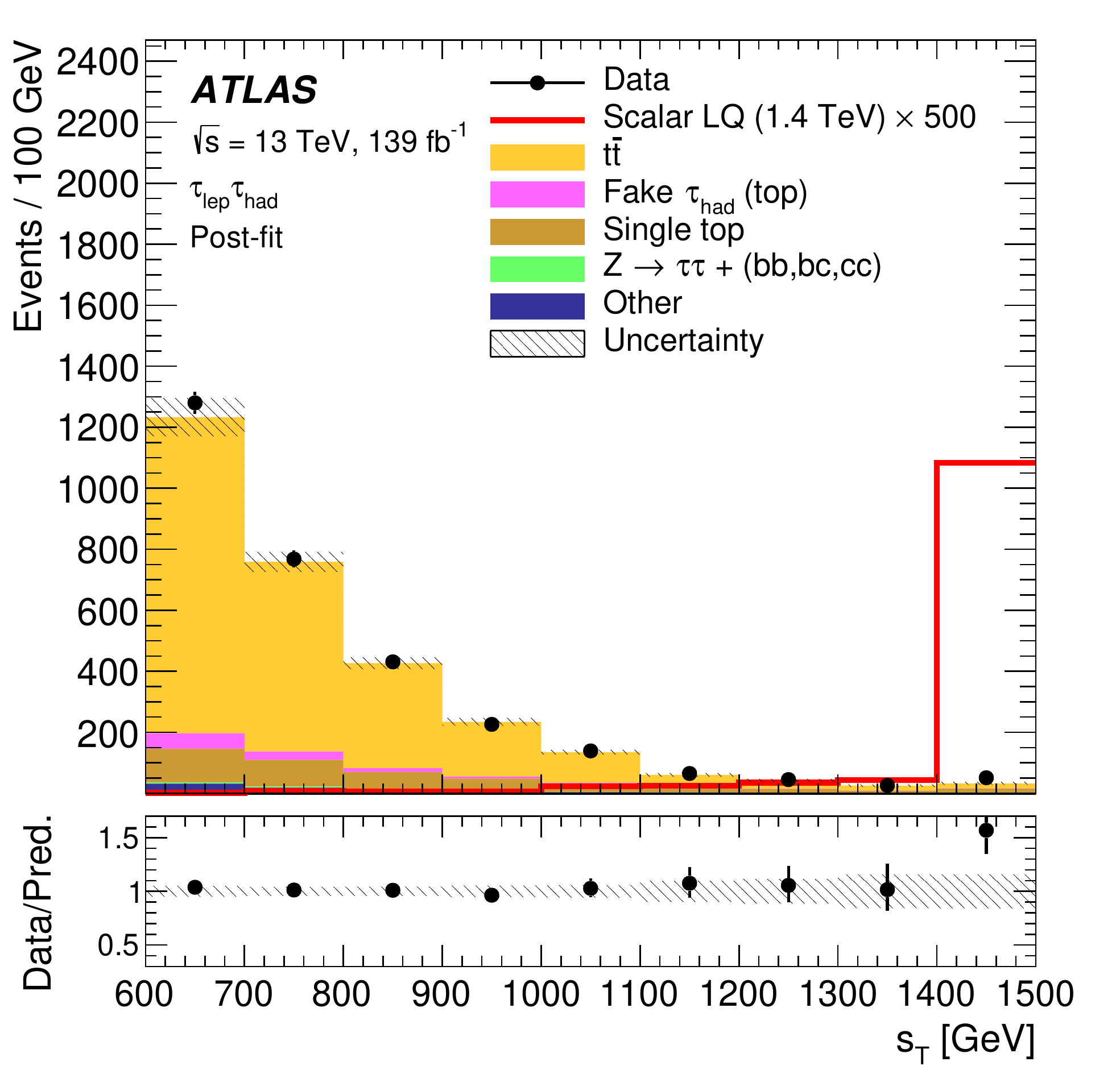}}
\caption{Signal (solid lines), post-fit background (filled histograms) and data (dots with
statistical error bars) distributions of representative
PNN input variables in the \lephad\ SR: (a) $\Delta R(\ell, \text{jet})$, (b),
$m(\tauhad, \text{jet})$ and (c) $\st$.  The normalisation and shape of the backgrounds are
determined from the background-only likelihood fit to data and the ratios of the data to the sum of
the predicted backgrounds are shown in the lower panels.  `Other' refers to the sum of minor backgrounds
(vector boson + jets, diboson and Higgs boson).  The hatched band indicates the combined statistical
and systematic uncertainty in the total background prediction.  The expected signal for a 1.4\,\TeV\
scalar LQ, scaled by the indicated factor for visibility, is overlaid. The last bin includes the overflow. }
 
\label{fig:mvainputs_LH}
\end{figure}

\begin{figure}
\centering
\subfigure[]{\includegraphics[width=.49\textwidth]{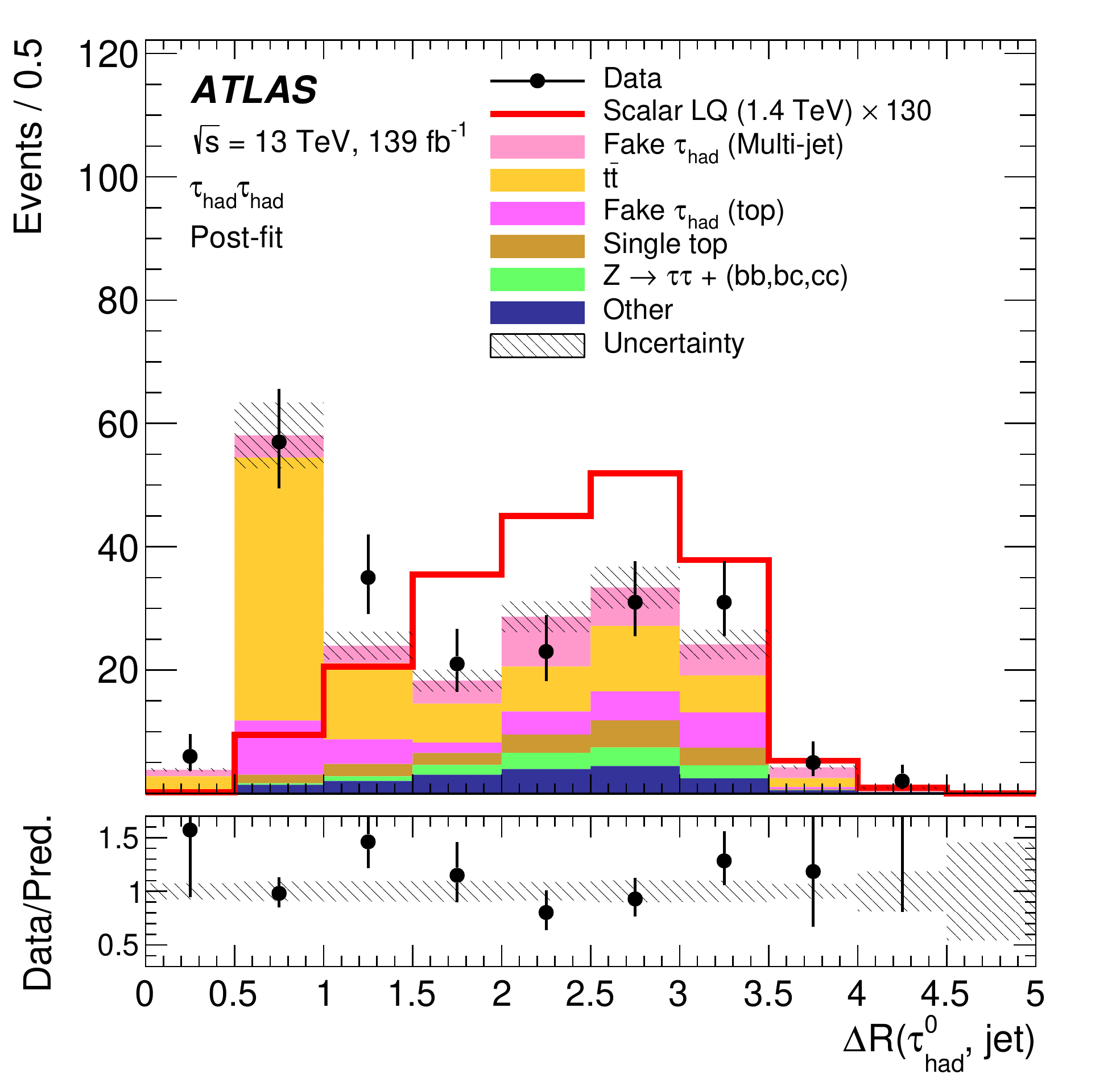}}
\subfigure[]{\includegraphics[width=.49\textwidth]{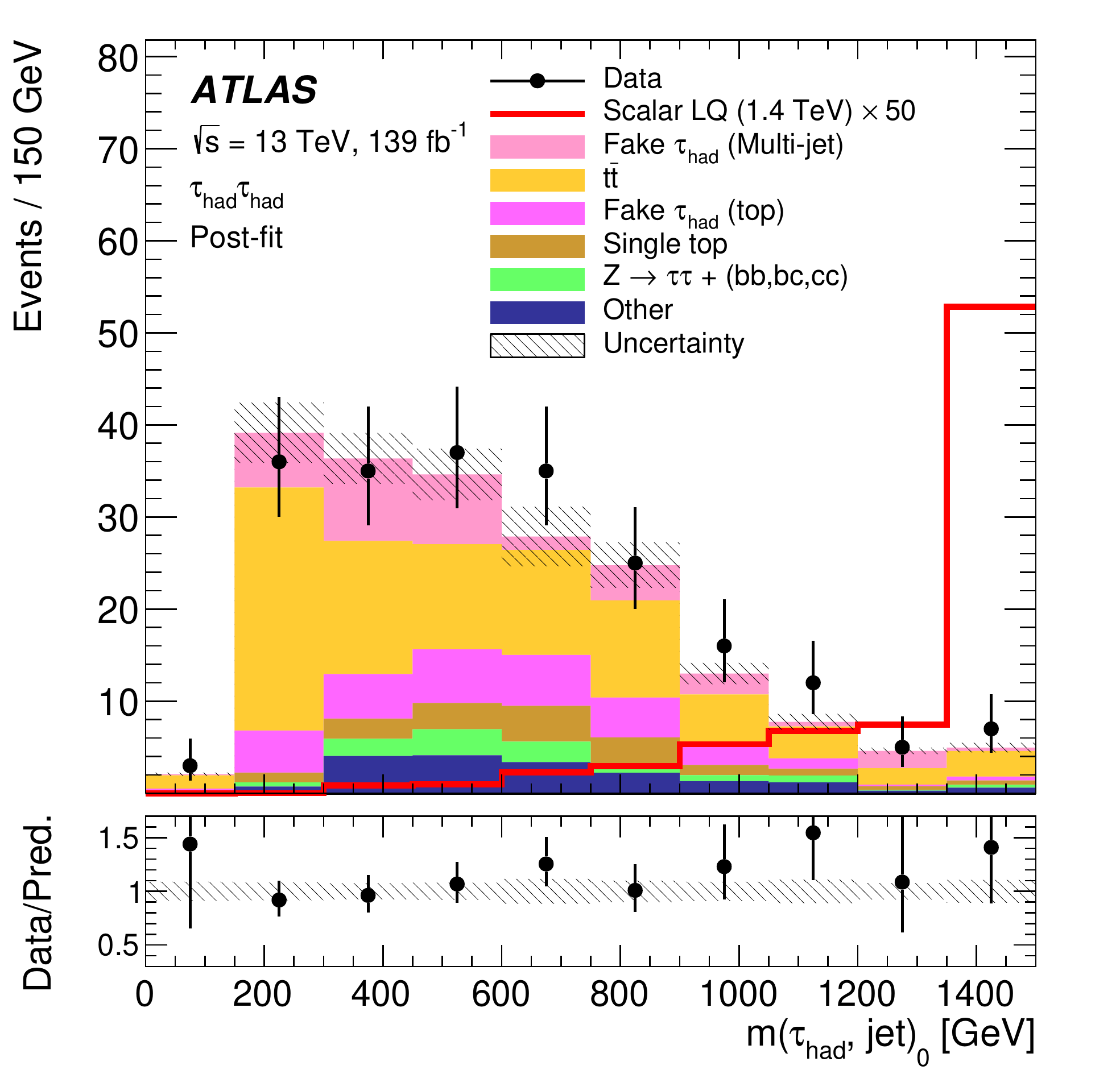}}
\subfigure[]{\includegraphics[width=.49\textwidth]{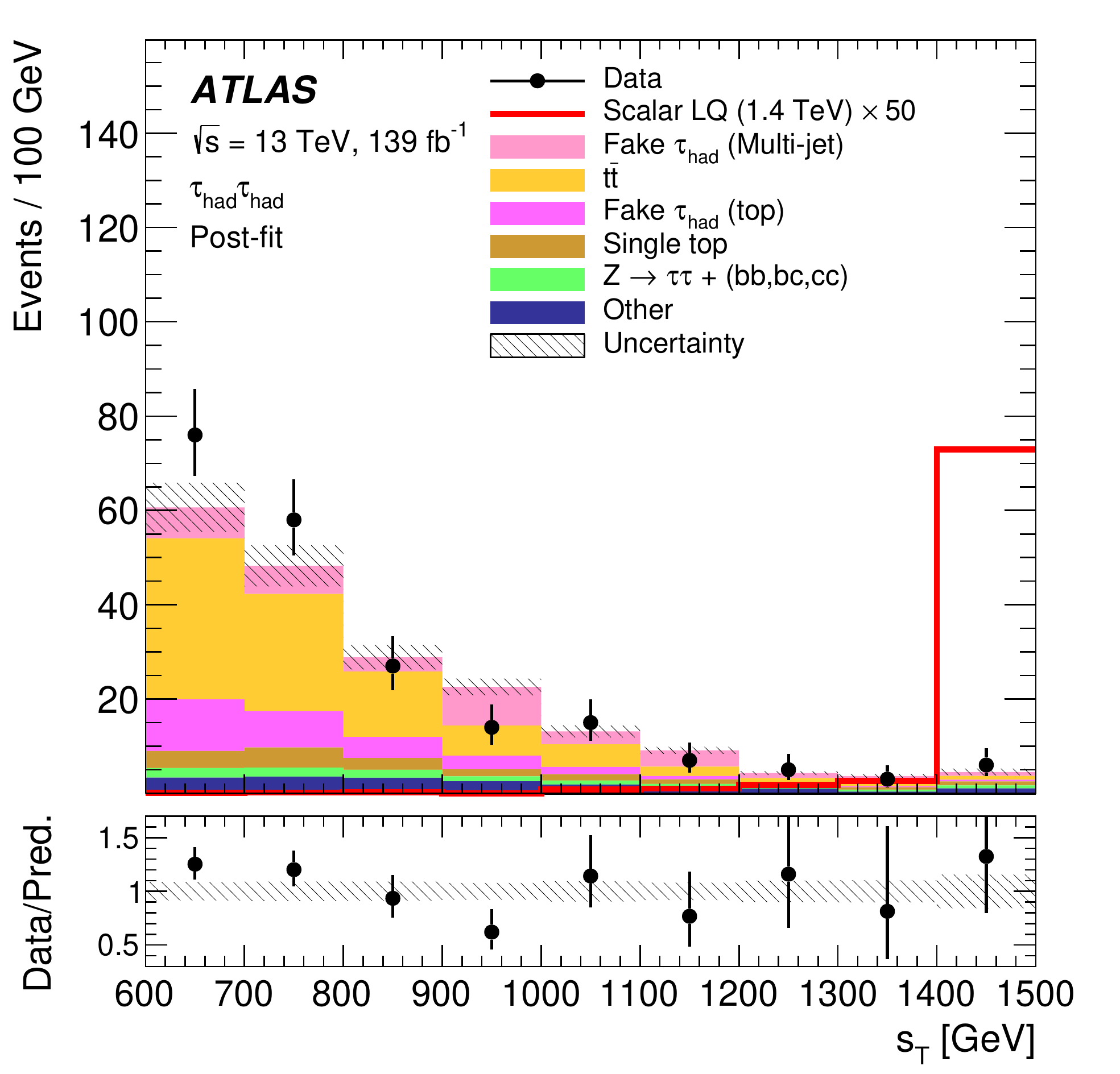}}
\caption{Signal (solid lines), post-fit background (filled histograms) and data (dots with
statistical error bars) distributions of representative
PNN input variables in the \hadhad\ SR: (a) $\Delta R(\tauhad^0, \text{jet})$ where $\tauhad^0$ is
the leading $\tau$-lepton, (b) the larger of the two $\tau$-jet mass combinations
$m(\tauhad, \text{jet})_0$ and (c) $\st$.  The normalisation and shape of the backgrounds are
determined from the background-only likelihood fit to data and the ratios of the data to the sum of
the predicted backgrounds are shown in the lower panels.  `Other' refers to the sum of minor backgrounds (vector boson + jets, diboson and
Higgs boson).  The hatched band indicates the combined statistical
and systematic uncertainty in the total background prediction.  The expected signal for a 1.4\,\TeV\
scalar LQ, scaled by the indicated factor for visibility, is overlaid.  The last bin includes the overflow.}
\label{fig:mvainputs_HH}
\end{figure}

A selection of representative input distributions, after the background corrections described
in~\Sect{\ref{sec:background}}, are presented in Figures~\ref{fig:mvainputs_LH}
and~\ref{fig:mvainputs_HH} for the \lephad\ SR and the \hadhad\ SR, respectively.  While the relative
importance of the variables varies with LQ mass, the $\st$ and mass variables are generally the most
performant.

\begin{table}[tbh]
\begin{center}
\caption{\protect\label{tab:mva}Summary of variables used as inputs to the PNN in the \lephad\ and
\hadhad\ categories.  The variables are defined in the text.}
\begin{tabular}{lcc}
\toprule
 
Variable & \lephad\ channel & \hadhad\ channel\\
 
\midrule
$\tauhadvis$ $\pt^{0}$ & \ding{51} & \ding{51}\\
$\st$  & \ding{51} & \ding{51}\\
$N_{b-\text{jets}}$ & \ding{51} & \ding{51} \\
$m({\tau, \text{jet}})_{0,1}$ & & \ding{51} \\
$m(\ell, \text{jet})$, $m(\tauhad, \text{jet})$ & \ding{51} \\
$\Delta R(\tau, \text{jet})$ & \ding{51} & \ding{51}\\
$\Delta \phi(\ell, \met)$ & \ding{51} & \\
$\met$ $\phi$ centrality & \ding{51} & \ding{51} \\
\bottomrule
 
\end{tabular}
\end{center}
\end{table}


\FloatBarrier
 
The PNNs are trained on all scalar LQ signal masses simultaneously against the main $t\bar{t}$ and
single-top backgrounds, taking into account both the true and misidentified \tauhadvis components
with the latter corrected as described in \Sect{\ref{sec:background}}.  The same PNN training is
used for both vector LQ models since separate trainings are found to provide a negligible
improvement in sensitivity.  For the signals, the generated LQ mass is used as the parameterisation
input in addition to the input variables described above, while in the case of the backgrounds a
mock LQ mass is randomly assigned from the range of signal LQ masses such that the resulting
training is independent of the mass. In all cases, the input variables are standardised by
subtracting the median value and dividing by the interquartile range.  Comparing the output between
the training dataset and an independent testing dataset showed no sign of overtraining.  The
resulting PNN score distributions, which peak at higher values for LQ signals than for the background
processes, are used as the final analysis discriminants.


\section{Background modelling}
\label{sec:background}

The dominant background in the $\hadhad$ and $\lephad$ channels is top production,
including $t\bar{t}$ and single-top-quark production. A subdominant background is $Z$ boson
production in association with heavy-flavour quarks $(bb, bc, cc)$, termed \zcr\ hereafter. Both
top production and \zcr\ are estimated from simulation to which data-driven corrections are applied.
In the $\hadhad$ channel, multi-jet events form a non-negligible background that is estimated by using
data-driven techniques. Small contributions to the background from all other processes are
estimated by using simulated events. This section describes the background estimation methods used
for top-quark-pair and single-top backgrounds, multi-jet backgrounds, and the \zcr\ background. The
background is validated for $\hadhad$ and $\lephad$ events in a region with an inverted $\st$ selection, as well as a
region with a low PNN score and the signal region selection. In addition, the $\hadhad$ multi-jet estimate
is validated in a region where the two $\tauhadvis$ have the same electric charge. The potential signal contamination in all regions
described in this section is negligible.
 
The process of estimating the backgrounds follows several steps.  First, an overall shape correction
is determined for the top background, as described in Section~\ref{sec:toprw}. Then, with this in
place, a shape and normalisation correction is determined for the top backgrounds with jets
misidentified as $\tauhadvis$, as described in Section~\ref{sec:topSF}. After applying these
corrections, a prediction for the shape and normalisation of multi-jet backgrounds is determined for
the $\hadhad$ channel in Section~\ref{sec:mj}.  Finally, with all relevant corrections in place, a
normalisation factor is determined for the \zcr\ backgrounds, as described in Section~\ref{sec:zcr}.
The resulting corrections are only weakly coupled due to the high purity of each control region,
meaning that corrections for a specific background process do not significantly affect the
overall background in control regions targeting other backgrounds. All of these corrections are applied in the final SR fit.
 
\subsection{Top quark backgrounds}
 
For top-quark-pair and single-top-quark production (top backgrounds), events are
estimated separately based on whether the $\tauhadvis$ candidate in the event is correctly identified (referred
to as a true $\tauhadvis$) or whether it is a quark- or gluon-initiated jet misidentified as a
$\tauhadvis$ (referred to as a fake $\tauhadvis$). The small contributions from light leptons that
are misidentified as $\tauhadvis$ are considered together with the true $\tauhadvis$
contribution. Events with a true $\tauhadvis$ and a hadronic jet misidentified as a light lepton
contribute negligibly to the $\lephad$ channel and are not considered.
 
These backgrounds are estimated in a multi-step data-driven process that is applied to simulated events. First, all top backgrounds are
scaled by an $\st$-dependent reweighting factor (RF), and then simulated background events with
misidentified $\tauhadvis$ are further corrected by a scale factor (SF) that is binned in
the \tauhadvis $\pt$.

\subsubsection{Overall reweighting of top backgrounds}
\label{sec:toprw}
 
The motivation for scaling the $t\bar{t}$ and single-top backgrounds arises from mismodelling of the data by simulation observed in control
regions (CRs).  It is seen that this effect becomes more pronounced for events with higher momentum top quarks, which is where this analysis
is primarily focused.  This mismodelling has also been observed in ATLAS measurements of the $\ttbar$ differential cross-section, where it is seen
that the number of events is overestimated at high top-quark $\pt$~\cite{TOPQ-2018-17,TOPQ-2018-18,TOPQ-2018-15}.
 
For this reason, a CR is defined to determine a binned shape and normalisation correction of the simulated top quark
events to data.  Events in this CR are required to have two $b$-jets with $\pt$ greater than 45 and
20~\GeV, exactly two light leptons ($ee$, $\mu\mu$ or $e\mu$) with opposite charge,
$\met > 100$ \,\GeV, and a dilepton mass ($m_{\ell \ell}$) $>110$ \,\GeV. They are also required to have $m_{b\ell} > 250$~\GeV, where $m_{b\ell} = \min(\max(m_{b_{0}\ell_{0}},m_{b_{1}\ell_{1}}),\max(m_{b_{0}\ell_{1}},m_{b_{1}\ell_{0}}))$,
where the 0 and 1 indices refer to the leading and sub-leading $b$-tagged jets and leptons in order of transverse momentum.
This region is orthogonal to the SRs and is over 99\% pure in $t\bar{t}$ events.
 
The RFs are derived by subtracting all non-top backgrounds, as estimated using simulation, from data. A ratio of the remaining events to the prediction of $\ttbar$
and single-top events in simulation is then calculated. This factor is binned
in $\st$, with one bin up to 400\,\GeV, steps of 100\,\GeV\ from 400 to 1400\,\GeV,
and then one bin for values greater than 1400\,\GeV. The values of the RFs decrease from 0.97 at low $\st$ to
approximately 0.62 in the highest $\st$ bin. Even in the highest $\st$ bin, the signal contamination remains at the percent level.
The largest relative contribution of single-top events is also at high $\st$. This reweighting is applied in both the $\lephad$ and $\hadhad$ SRs for $t\bar{t}$
and single-top events with true and misidentified $\tau$-leptons, as well as in all CRs. The uncertainty in this RF is
taken from the statistical  uncertainty in the factor, bin-by-bin in $\st$, and its impact on the shape
and normalisation of the final PNN score distribution are considered.
In addition, top background modelling uncertainties are propagated through the reweighting process, so that modified RFs are applied when evaluating such uncertainties in the final fit.
 
\subsubsection{Top backgrounds with jets misidentified as $\tauhadvis$}
\label{sec:topSF}
 
In addition to this overall RF, the estimation of top backgrounds with jets misidentified as $\tauhadvis$ in the SRs is
performed using simulated events with additional data-driven corrections. A fit is performed in a $\lephad$-based CR to simultaneously correct the
overall normalisation of true $\tauhadvis$ and misidentified $\tauhadvis$ events while deriving an SF to be applied to misidentified $\tauhadvis$ events in the
$\lephad$ and $\hadhad$ SRs. The RF for top backgrounds is applied to this CR before the fit. The  SFs obtained are then applied in the SRs, in order
to correct the $\tauhadvis$ misidentification rate in simulation to that observed in data.
 
The CR has the same selection as the SR
for the $\lephad$ channel, except that the
$\tauhadvis$ $\pt > 100$~\GeV\ requirement is removed and $\st$ is required to be in a range of
400--600\,\GeV. This region is 97\% pure in $\ttbar$ events, with a mixture of both correctly identified and
misidentified $\tauhadvis$ that varies with $\tauhadvis$ $\pt$.
 
\begin{figure}[htp]
\centering
\includegraphics[width=0.45\textwidth]{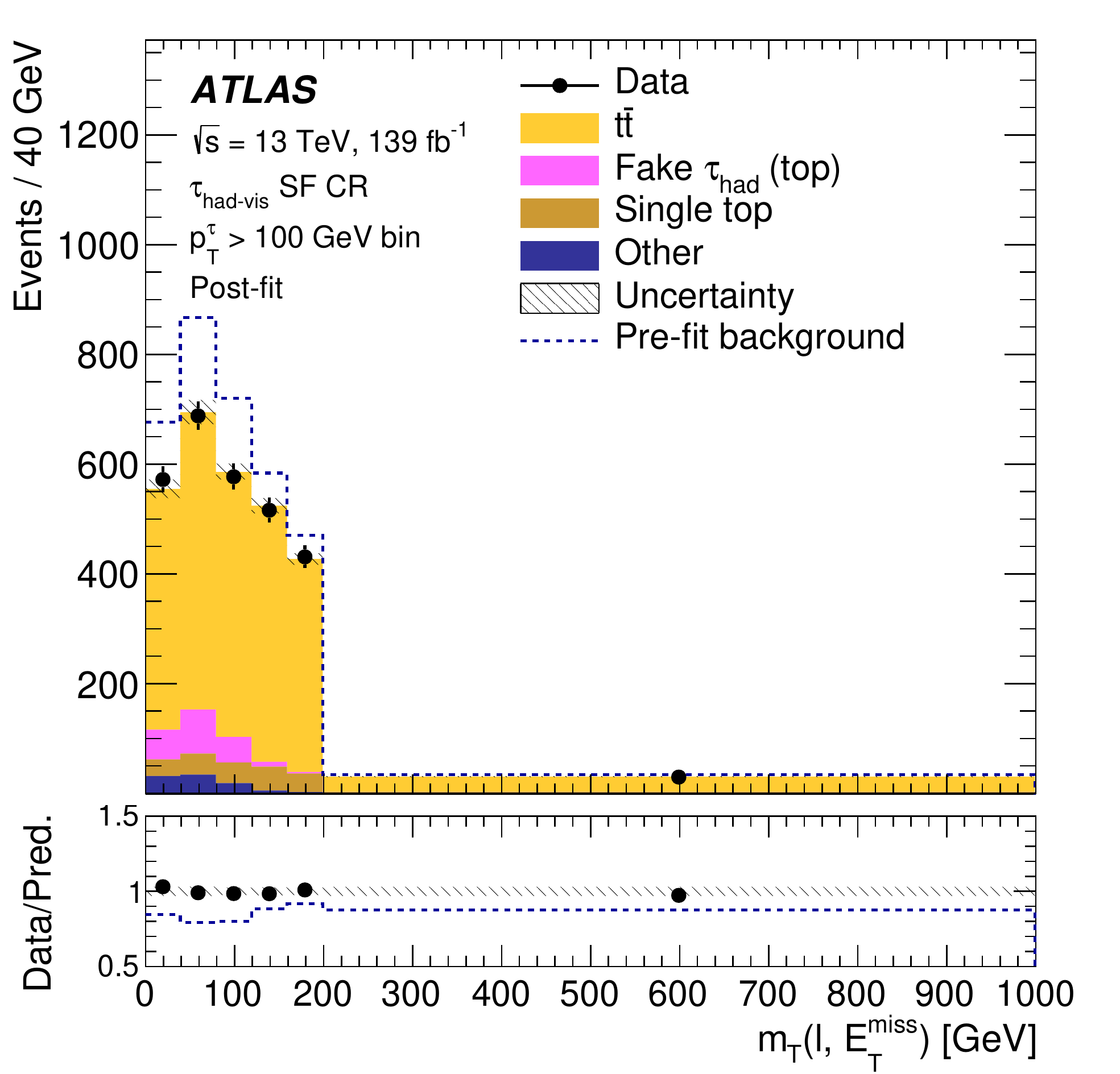}
\caption{Post-fit plots for true and misidentified $\tauhadvis$ in the $\tauhadvis$
CR, in an example $\pt$ bin ($\tauhadvis$ $\pt$ $>$100~\GeV).  `Other'
refers to the sum of minor backgrounds (vector boson + jets, diboson and Higgs boson). The lower panels show the ratios of the data to the sum of the predicted backgrounds. The
hatched bands indicate the combined statistical and systematic uncertainty in the total
background predictions. The dashed lines denote the total pre-fit backgrounds for comparison,
while the last bins include the
overflow.}  \label{fig:SF_CRs}
\end{figure}

The distribution used for this estimation is the transverse mass of the light lepton and missing
transverse momentum, defined as
$m_{\text{T}}(\ell, \met) = \sqrt{(\met + p_{\text{T},\ell})^2 - (E_{\text{T},x}^{\text{miss}} + p_{x,\ell})^2 - (E_{\text{T},y}^{\text{miss}} + p_{y,\ell})^2 }$.
The expected shapes for top backgrounds with true and misidentified
$\tauhadvis$ in this distribution differ significantly, making it possible to constrain the two
background sources.  The normalisation of the true and misidentified $\tauhadvis$ background is
allowed to vary freely, and SFs for the misidentified $\tauhadvis$ background are determined in bins
of $\tauhadvis$ $\pt$.  All detector-related uncertainties and top background modelling
uncertainties are included as nuisance parameters in the fit. An example fit in a single bin of
$\pt$ is shown in Figure~\ref{fig:SF_CRs} for the $\tauhadvis$ CRs. Depending on
$\tauhadvis$ $\pt$, the SFs run from 0.90 in the lowest $\pt$ bin down to $0.56$ in the highest
$\pt$ bin.
 
For the estimation of top backgrounds with misidentified $\tauhadvis$, an uncertainty is considered that arises  from
the limited number of events and an additional uncertainty is defined by comparing the nominal SFs to SFs
derived with a more inclusive $\st$ selection ($\st<600$~\GeV). This last uncertainty is intended to address a possible $\st$-dependence in
the mismodelling of top backgrounds. The difference between the central values for SFs measured with these two $\st$ selections
is taken as the $\st$-dependence uncertainty.
 
\subsection{Multi-jet backgrounds with jets misidentified as $\tauhadvis$}
\label{sec:mj}

For the $\hadhad$ channel, multi-jet processes can contribute to the SR at non-negligible levels.
For this reason, the $\hadhad$ channel uses a data-driven fake-factor (FF) method to estimate this
background. These FFs are measured in a CR with the same selection as the \hadhad\ SR,
except that the two $\tauhadvis$ candidates have the same charge and the $\met$ requirement is
loosened to 80~\GeV. The FF is defined as the ratio of events where both $\tauhadvis$ are loose to
the number of events where one $\tauhadvis$ is loose and the other is an anti-$\tauhadvis$. These
FFs are derived as a function of transverse momentum and the number of charged-particle tracks of the
$\tauhadvis$ candidate. The FFs are measured from data after subtracting all predicted non-multi-jet
background contributions. The FFs range between approximately zero and 0.25.
 
\begin{figure}[htp]
\centering
\subfigure[]{\includegraphics[width=0.45\textwidth]{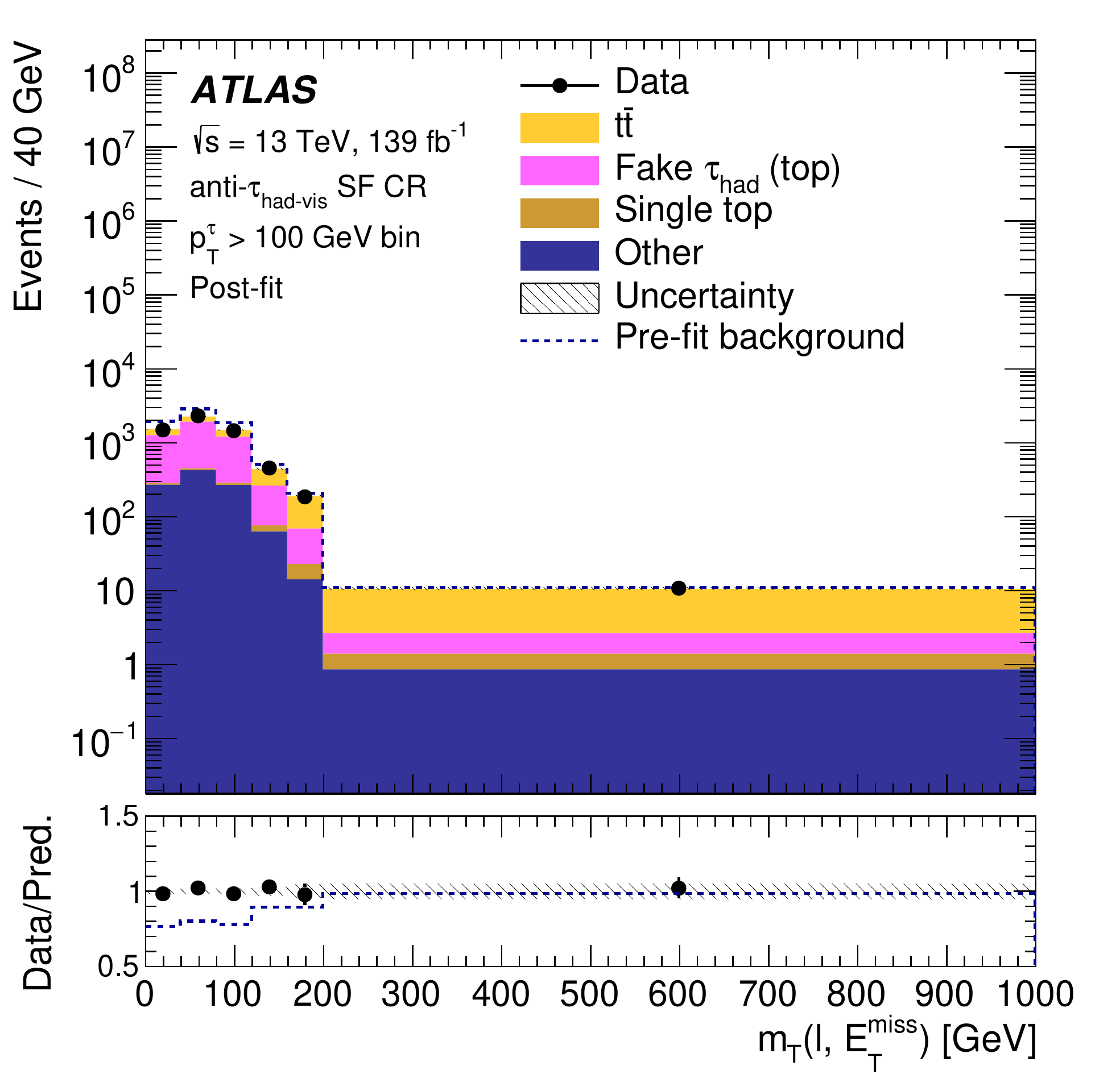}}
\caption{Post-fit plot for true and misidentified $\tauhadvis$ in the
the anti-$\tauhadvis$ CR, in an example $\pt$ bin ($\tauhadvis$ $\pt$ $>$100~\GeV).  `Other'
refers to the sum of minor backgrounds (vector boson + jets, diboson and Higgs boson). The lower panels show the ratios of the data to the sum of the predicted backgrounds. The
hatched bands indicate the combined statistical and systematic uncertainty in the total
background predictions. The dashed lines denote the total pre-fit backgrounds for comparison,
while the last bins include the
overflow.}  \label{fig:SF_CRs_antitau}
\end{figure}

The non-multi-jet background contributions that are subtracted, however, suffer from the same
mismodelling issues described in the previous two sections.  The top backgrounds are therefore
corrected by the RFs and SFs derived as described in Section~\ref{sec:toprw} and
Section~\ref{sec:topSF}, respectively. Since the SFs are anticipated to be different for fake $\tauhadvis$
passing the $\tauhadvis$ and anti-$\tauhadvis$ requirements, dedicated SFs are measured for this data-driven
estimation. Specifically, the anti-$\tauhadvis$ region uses SFs that are derived in a CR
as described in Section~\ref{sec:topSF}, except that the $\tauhadvis$ identification requirement is
changed to that of an anti-$\tauhadvis$.  An example fit in a single bin of $\pt$ is shown in
Figure~\ref{fig:SF_CRs_antitau} for the anti-$\tauhadvis$ CR. Depending on $\tauhadvis$ $\pt$,
these SFs vary between 0.77 and 0.95.
 
To construct the background estimate, FFs are applied to a region with the \hadhad\ SR selection,
except that the $\tauhadvis$ identification requirement is changed to that of an anti-$\tauhadvis$.
This provides both a shape and a normalisation for the multi-jet contribution in the PNN score
distribution.
 
For the estimation of multi-jet backgrounds in $\hadhad$, uncertainties are considered due to the
statistical uncertainty of the FFs, decorrelated in $\pt$ bins, and to the uncertainty in the subtraction of different
backgrounds using simulation. Top events with a correctly identified $\tauhadvis$, top events with a
misidentified $\tauhadvis$, and other small backgrounds are considered separately. The top events
are varied by the overall uncertainty defined by the procedure to determine the modelling
uncertainties, but evaluated in the anti-$\tauhadvis$ region, resulting in an uncertainty of
approximately 50\%.  The other backgrounds are varied by a conservative value of 30\%; inflating the
size of this uncertainty further was found to have negligible impact on the result.  In addition, a
20\% overall uncertainty in the estimate is applied based on checks of the method in the validation
regions described above. The total uncertainty in the multi-jet background is $-64$\% and $+61$\%.
 
\subsection{\zcr\ background}
\label{sec:zcr}
 
The normalisation of the \zcr\ background, which is a relatively small contribution in the SRs, is
observed to be in disagreement with the NLO cross-section in \textsc{Sherpa} (e.g.\
Ref.~\cite{HIGG-2016-29}). It is therefore determined from data using a \zcr\ CR that targets events
containing a $Z$ boson decaying into a light-lepton pair and produced in association with two
heavy-flavour jets. The composition of this control region is approximately 60\% \zcr\ events and 40\%
$\ttbar$ events, with less than 1\% arising from backgrounds with misidentified $\tauhadvis$. Since the
contribution from backgrounds with misidentified $\tauhadvis$ is negligible, only the RF
for the $\ttbar$ shape is included in this CR.
 
Data for the CR was recorded using a combination of the single-lepton triggers described
above and additional dilepton triggers requiring pairs of same-flavour leptons.  At the
analysis level exactly two oppositely-charged same-flavour leptons, passing the `veto' quality
requirements and \pt\ thresholds based on the corresponding trigger thresholds, are required.
The invariant mass of the resulting lepton pair \mll\ is required to lie between
75\,\GeV\ and 110\,\GeV.  In addition, exactly two $b$-jets with $\pt > 20$\,\GeV\ are required
and their invariant mass $m_{bb}$ is required to be less than 40\,\GeV\ or greater than 150\,\GeV\ to avoid the
Higgs boson mass peak. The RFs for top backgrounds derived in Section~\ref{sec:toprw} are
then applied.
 
A fit to the \mll\ distribution is performed to discriminate between the \zcr\ and top
backgrounds, with the normalisation of both processes allowed to vary freely
and all systematic uncertainties described in \Sect{\ref{sec:systs}} included.  The
resulting \zcr\ normalisation factor is $1.36 \pm 0.11$ and is used to correct the \zcr\ background entering into
the final fit (described in \Sect{\ref{sec:result}}), which is allowed to vary within the
associated uncertainty.


\section{Systematic uncertainties}
\label{sec:systs}

The systematic uncertainties considered include detector-related uncertainties, modelling and theoretical uncertainties,
and uncertainties derived for the data-driven background estimates, the latter of which have
already been described in Section~\ref{sec:background}. Uncertainties are evaluated by shifting the central value upward or
downward by one standard deviation, and then propagating the differences to
the PNN score distributions that are used in the final fit.
 
Detector-related uncertainties are defined as uncertainties relating to the detector response,
object reconstruction and object identification.  There are systematic uncertainties associated with
each of the reconstructed objects considered, as well as the $\met$. For light leptons,
$\tauhadvis$, and jets, uncertainties are considered for energy scale and resolution, reconstruction and
identification, while uncertainties in isolation are also considered for light leptons.
For the $\lephad$ and $\hadhad$ channel, uncertainties associated with the lepton and $\tauhadvis$
trigger efficiencies, respectively, are considered. For $b$-jets, additional uncertainties are considered
for the efficiency of (mis)tagging $b$-jets, $c$-jets, and light-quark-initiated jets.
Uncertainties related to energy scale and resolution, and the inclusion of soft terms,
are considered for the $\met$.  Finally, there is also an
uncertainty associated with shape and normalisation components that arises from uncertainties in the simulation
of pile-up collisions.

Theoretical and modelling uncertainties include uncertainties in the cross-section calculations, which
have only a normalisation component, and uncertainties in the acceptance, for which
normalisation and shape components are considered. For top backgrounds, relative acceptance uncertainties are
also defined to take into account normalisation differences for \lephad\ and \hadhad\ SRs.
 
For $\ttbar$ processes, shape and normalisation uncertainties are considered that arise from
changing the matrix element and parton shower simulation software, and from varying the initial and final state
radiation, PDF, and $\alpha_s$. The matrix element uncertainty is determined by comparing
the \textsc{Powheg}+\textsc{Pythia 8} sample with an \textsc{aMC@NLO}+\textsc{Pythia 8} sample. The parton shower uncertainty is
determined by comparing the \textsc{Powheg}+\textsc{Pythia 8} sample with a \textsc{Powheg}+\textsc{Herwig 7}~\cite{Bahr:2008pv,Bellm:2015jjp} sample. The other modelling
uncertainties are evaluated using internal weights in the nominal $\ttbar$ sample.
 
For single-top processes, acceptance uncertainties with shape and normalisation components are considered.
Uncertainties are considered from changing the matrix element, parton shower, and impacts of diagram
interference. In addition, variations of initial- and final-state radiation and PDFs are considered. The matrix element uncertainty is determined by comparing
the \textsc{Powheg}+\textsc{Pythia 8} sample with an \textsc{aMC@NLO}+\textsc{Pythia 8} sample. The parton shower uncertainty is
determined by comparing the \textsc{Powheg}+\textsc{Pythia 8} sample with a \textsc{Powheg}+\textsc{Herwig 7} sample.
The diagram interference uncertainty is evaluated by comparing the nominal single top samples, which use a diagram removal scheme,
with alternative samples that utilise a diagram subtraction scheme~\cite{Frixione:2008yi}.  The other modelling
uncertainties are evaluated using internal weights in the nominal single-top samples.
 
All $\ttbar$ and single-top modelling uncertainties are also propagated through the top reweighting  procedure, such that there is an
uncertainty in the RF corresponding to each modelling uncertainty.
 
For all $Z$+jets processes, uncertainties due to the choice of generator are evaluated by comparing the nominal \SHERPA
simulated samples with alternative samples simulated by \MADGRAPH with LO-accurate matrix elements that contain up to four final-state partons,
using \PYTHIA for parton showering. In addition, uncertainties are considered by
taking an envelope of variations in the renormalisation and factorisation scales and PDF values using internal weights in
the simulated \SHERPA sample. For this process specifically, uncertainties are also included based on varying the matrix element matching scale and the
resummation scale for soft-gluon emission. Uncertainties are considered separately for the \zcr\ background and the remaining $Z+(bl,cl,ll)$ background,
where $l$ indicates a light-flavour jet, and only normalization is taken into account. All of these uncertainties are included in the \zcr\ fit described in Section~\ref{sec:zcr}, and their sum
in quadrature, taking relative acceptance uncertainties into account, is considered as the uncertainty in the SRs for the final fit.
 
For other minor backgrounds, the following uncertainties are considered. A conservative uncertainty
of 50\% on the normalisation of the diboson backgrounds is included.  For $W$+jets processes, which
form a background to the analysis due to a jet being misidentified as a light lepton or a $\tauhad$,
a conservative 100\% uncertainty is taken into account, decorrelated between the $\lephad$ and
$\hadhad$ channels.  Within the $\lephad$ channel, the uncertainty is also decorrelated between the
cases where it is the light lepton or the $\tauhad$ that is misidentified.  In both cases, the choice
of the size of these uncertainties has a negligible impact on the results.
 
For signal samples, uncertainties arising from variations of scale, initial-state radiation, PDF, and
$\alpha_s$ are considered, using alternative weights internal to the signal samples. Differences in
shape are observed to be negligibly small in the PNN score distributions,
so only variations in normalisation are included for the final fit.

The relative impact of the different sources of uncertainty on the analysis varies depending on the
LQ model considered and the mass probed.  Generally, the largest impact comes from the statistical
uncertainties, which increase with \mlq.  In the scalar LQ case, for example, the statistical impact
on the limit ranges from 60\% at the lowest \mlq\ evaluated to 80\% above 1000\,GeV.  The main
systematic uncertainties come from the $\ttbar$ and single-top-quark modelling uncertainties,
including their interference, and normalisation.  There is also a significant effect from the signal
acceptance uncertainties, which increases with \mlq, particularly for the vector LQ models.


\section{Statistical interpretation and results}
\label{sec:result}

The data are compared with the expectation, including the background modelling corrections outlined
in \Sect{\ref{sec:background}}, by performing simultaneous binned maximum-likelihood fits to the PNN
score distributions, separately for each LQ hypothesis, in the \lephad\ and \hadhad\ SRs.  For each
hypothesis, the binning of the PNN score distributions is chosen separately to maximise the expected
sensitivity, while ensuring sufficient background events in the signal-enhanced PNN bins and preserving
the stability of the fit.  In addition to the relative signal-strength modifier, $\mu$, the top
normalisation is free to float in the fit and is constrained by the background-enhanced PNN bins.
 
The statistical and systematic uncertainties affecting the signal and background model, described
in \Sect{\ref{sec:systs}}, are represented by deviations from the nominal model scaled by Gaussian-
or Poisson-constrained nuisance parameters that are profiled in the fit.  Common sources of
systematic uncertainty are correlated across the SRs.
 
\begin{figure}
\centering
\subfigure[]{\includegraphics[width=.4\textwidth]{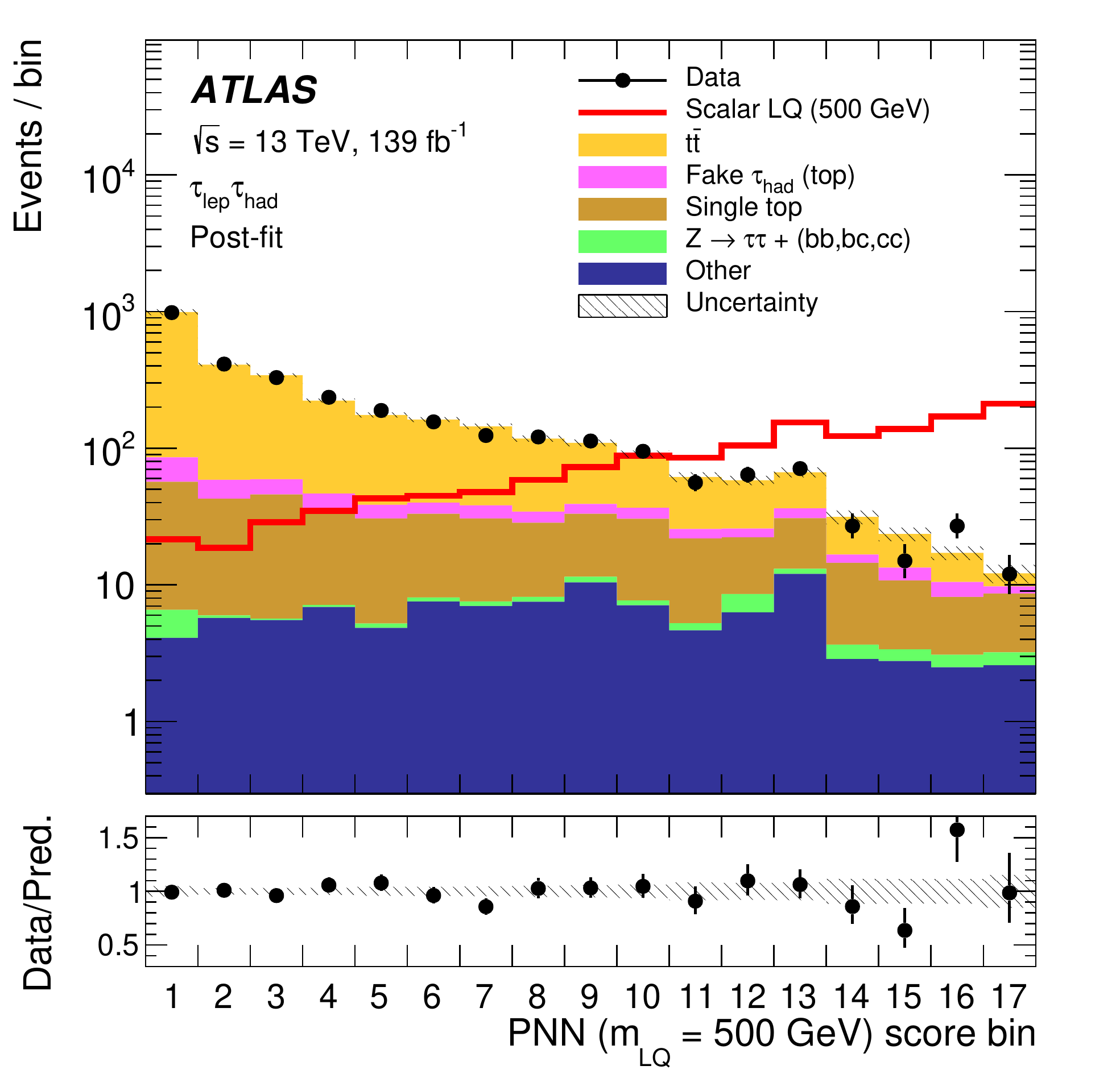}}
\subfigure[]{\includegraphics[width=.4\textwidth]{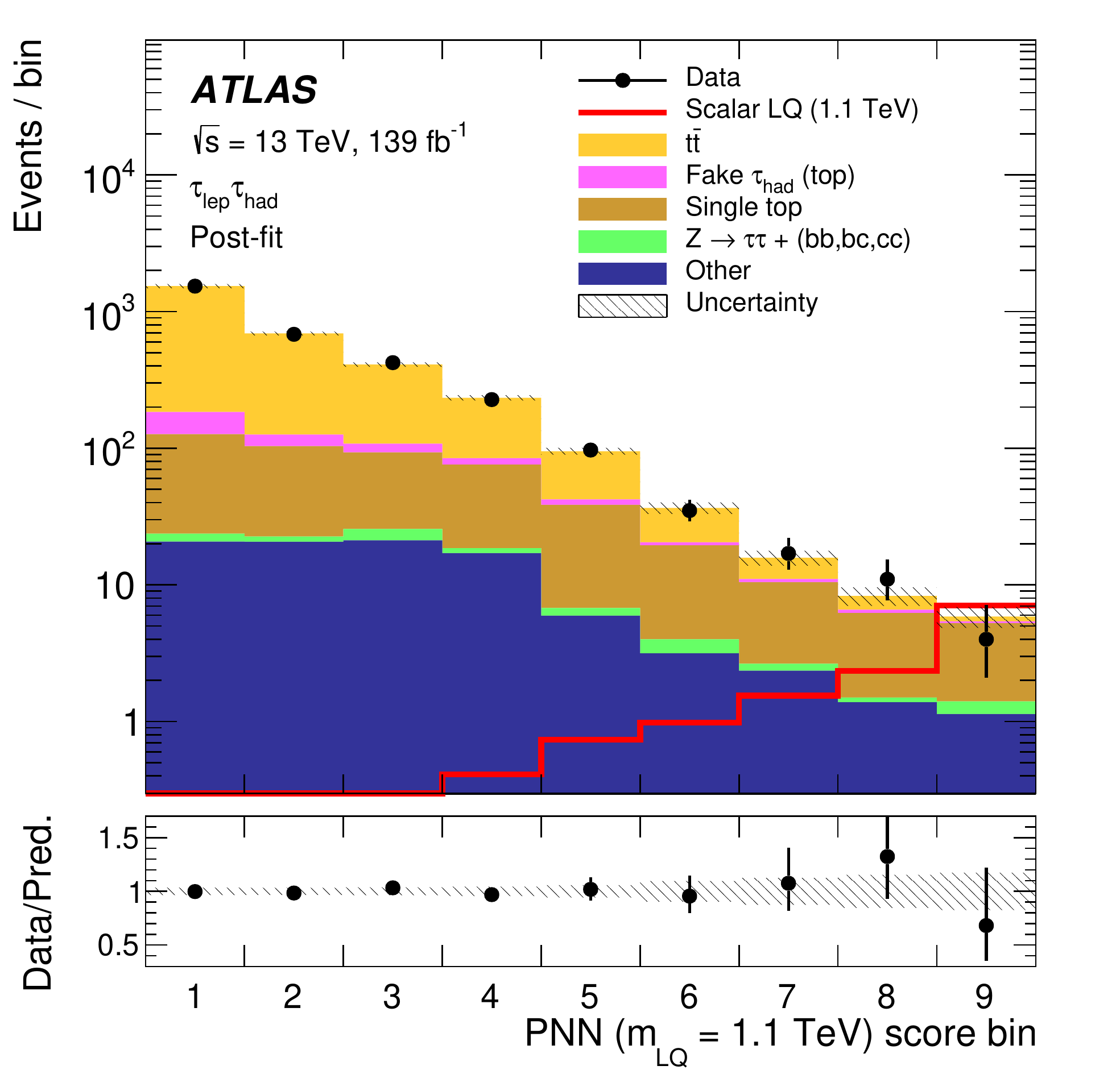}}
\subfigure[]{\includegraphics[width=.4\textwidth]{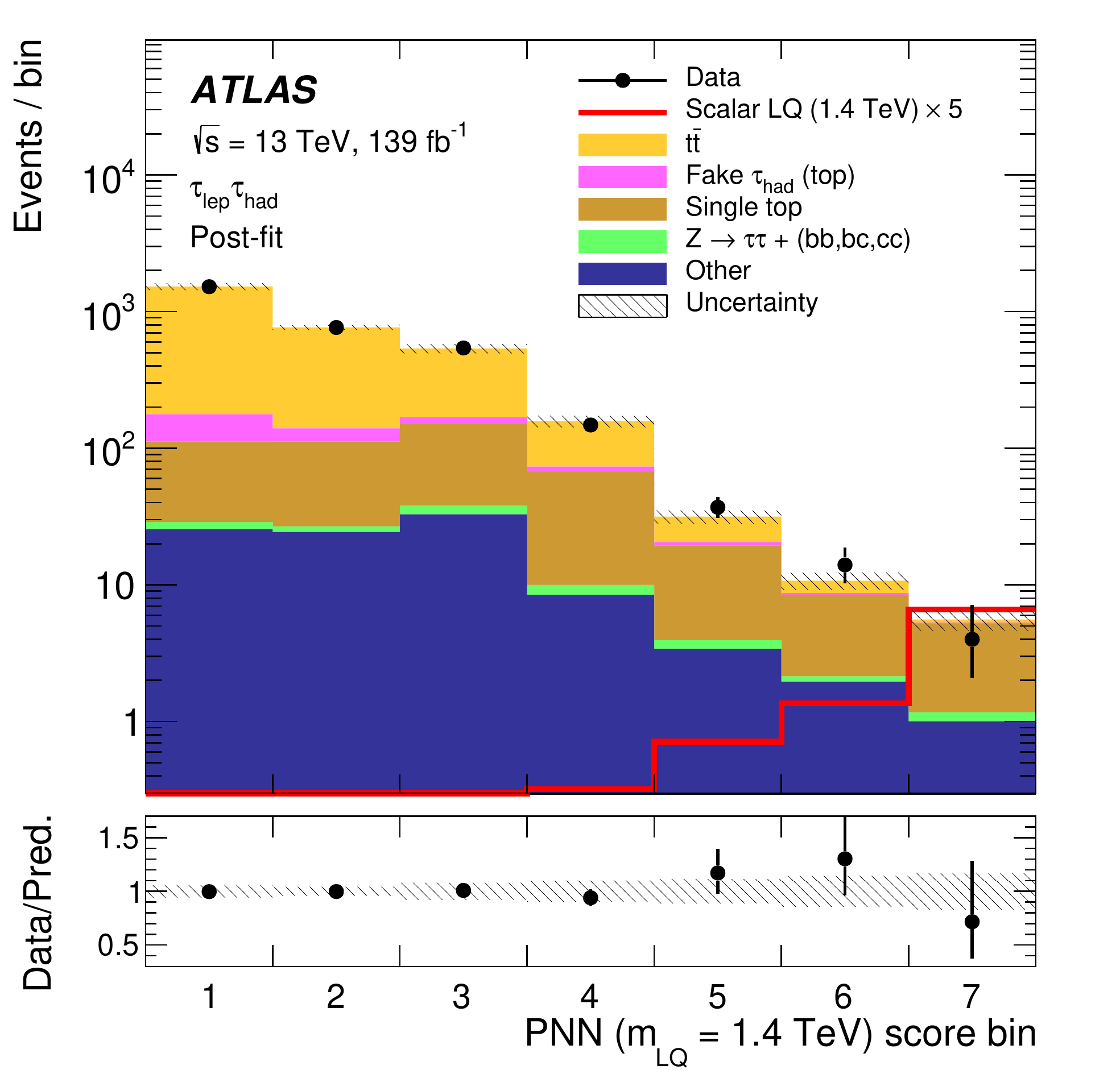}}
\caption{ The PNN score distributions in the \lephad\ SR for (a) $\mlq =
500$\,\GeV, (b) $\mlq = 1.1$\,\TeV, (c) $\mlq = 1.4$\,\TeV.  The normalisation and shape of the
backgrounds are determined from the background-only likelihood fit to data and the ratios of the data
to the sum of the backgrounds are shown in the lower panels.  `Other' refers to the sum of minor
backgrounds (vector boson + jets, diboson and Higgs boson).  The hatched bands indicate the combined
statistical and systematic uncertainty in the total background predictions.  The expected signals for
scalar LQs with the corresponding masses, scaled by the indicated factors for visibility, are overlaid.
Since the PNN score itself is not a physical quantity, it is represented solely by the bin
number.}
\label{fig:pnn_lh}
\end{figure}

\begin{figure}
\centering
\subfigure[]{\includegraphics[width=.4\textwidth]{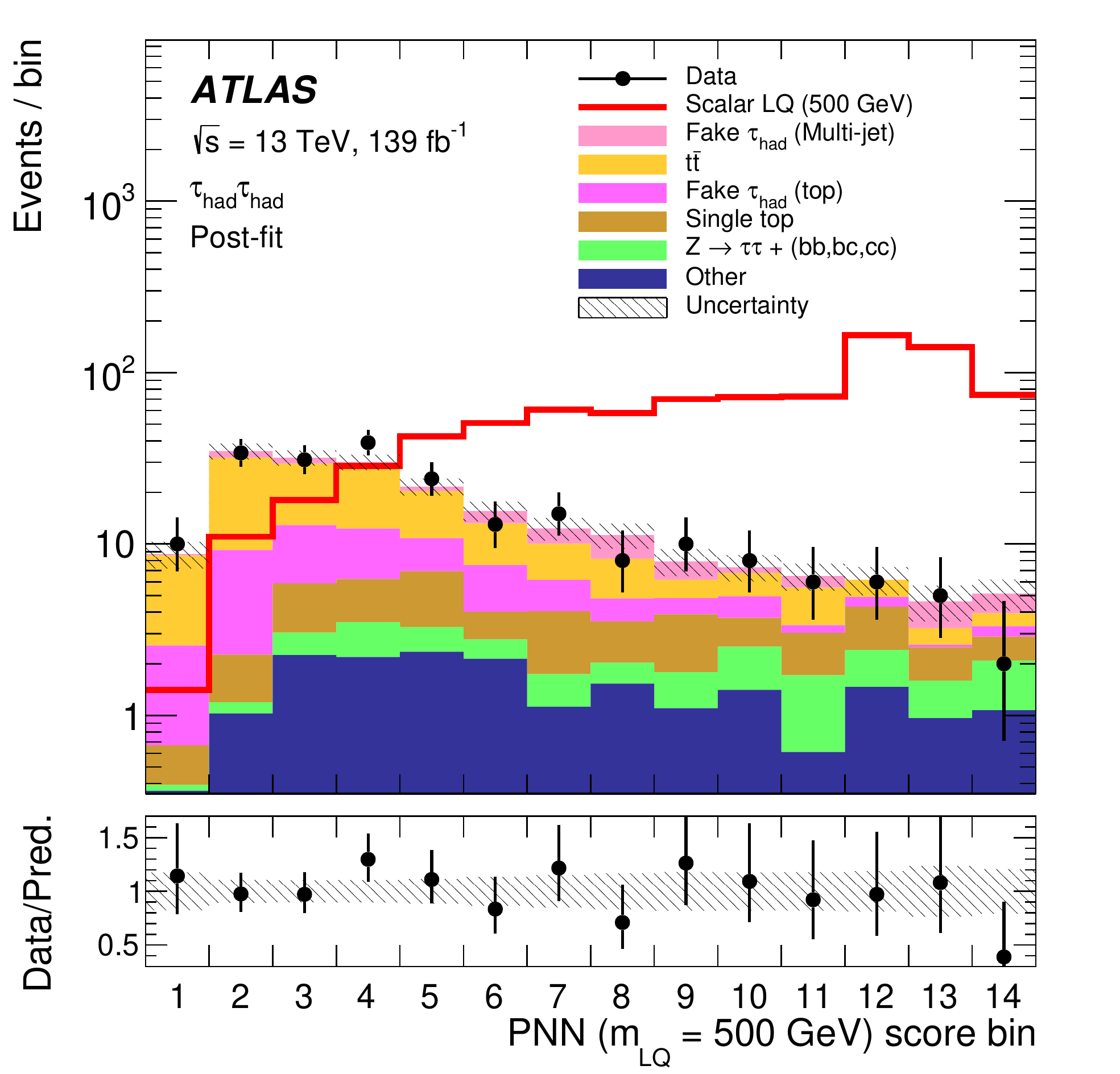}}
\subfigure[]{\includegraphics[width=.4\textwidth]{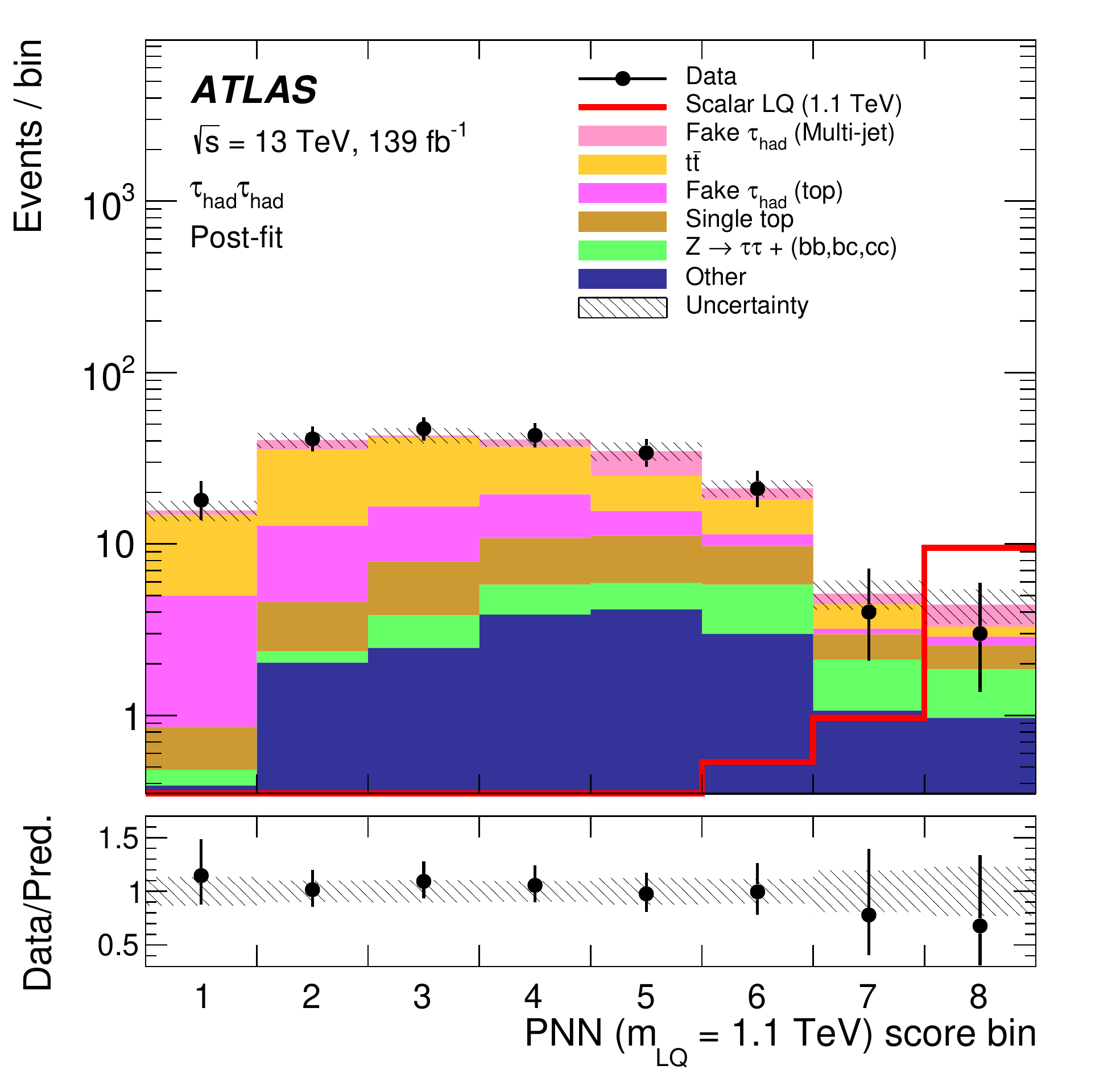}}\\
\subfigure[]{\includegraphics[width=.4\textwidth]{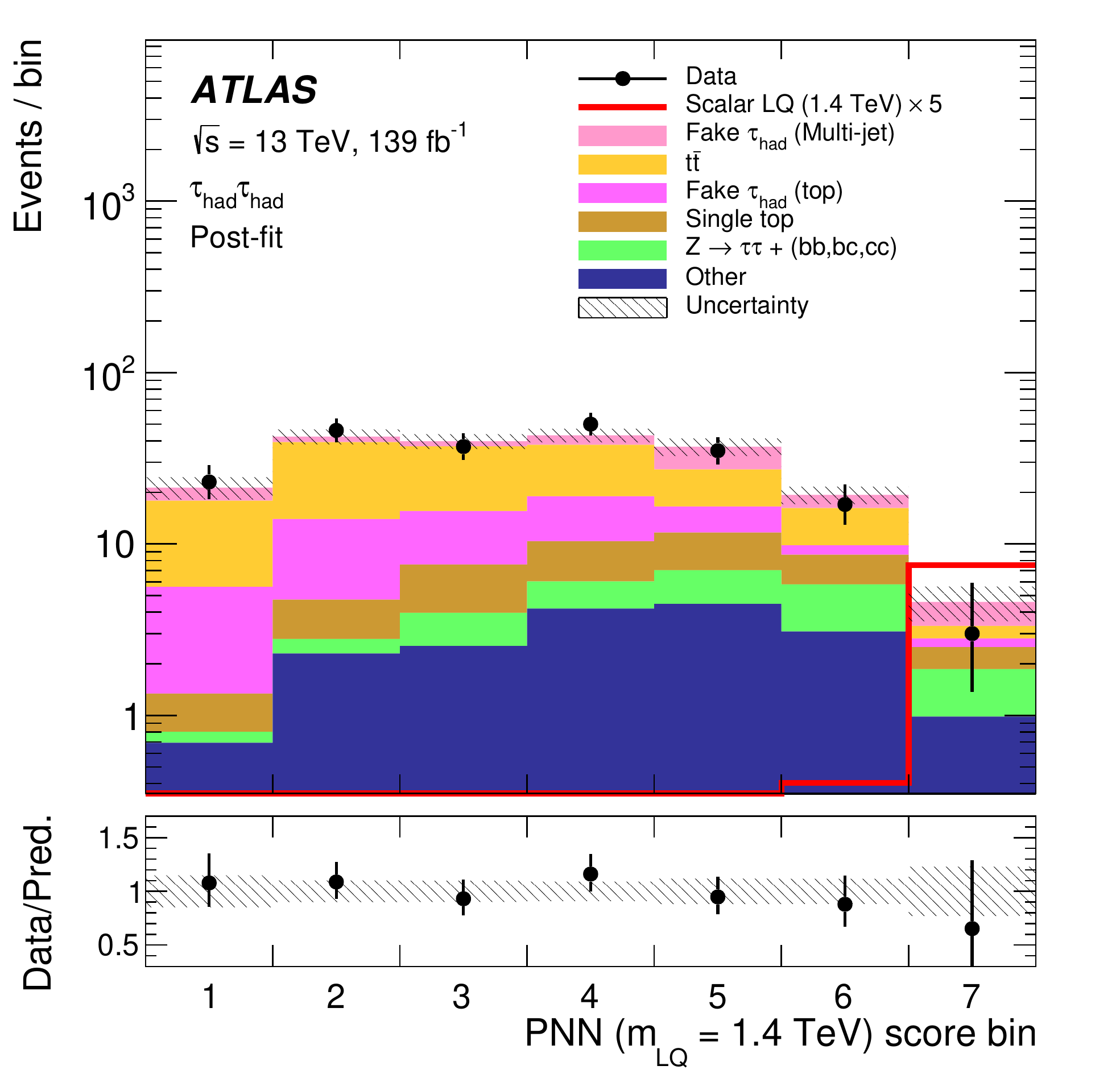}}\\
\caption{ The PNN score distributions in the \hadhad\ SR for (a) $\mlq =
500$\,\GeV, (b) $\mlq = 1.1$\,\TeV, (c) $\mlq = 1.4$\,\TeV.  The normalisation and shape of the
backgrounds are determined from the background-only likelihood fit to data and the ratios of the data
to the sum of the backgrounds are shown in the lower panels.  `Other' refers to the sum of minor
backgrounds (vector boson + jets, diboson and Higgs boson).  The hatched bands indicate the combined
statistical and systematic uncertainty in the total background predictions.  The expected signals for
scalar LQs with the corresponding masses, scaled by the indicated factors for visibility, are overlaid.
Since the PNN score itself is not a physical quantity, it is represented solely by the bin
number.}
 
\label{fig:pnn_hh}
\end{figure}

The resulting event yields in the \lephad\ and \hadhad\ SRs, based on a background-only fit to the
data, are presented in \Tab{\ref{tab:yields}}.
Corresponding post-fit PNN score distributions for representative LQ signals at masses of 500\,\GeV,
1.1\,\TeV\ and 1.4\,\TeV\ are shown in \Fig{\ref{fig:pnn_lh}} (\Fig{\ref{fig:pnn_hh}}) for the \lephad\
(\hadhad) SR\@.  At high values of the PNN score, top backgrounds dominate in the \lephad\
channel, while the \hadhad\ background consists of a roughly even mixture of all background sources.
Overall, good agreement with the SM background expectation is observed in all cases, although
there is a slight deficit of data relative to the background prediction in the highest PNN score
bin for the $\hadhad$ channel.

\begin{table}[!h]
\centering
\caption{Post-fit yields for background events, determined from a background-only fit, compared with
the observed number of data events in the \lephad\ and \hadhad\ SRs. `Fake \tauhad (top)' refers to
top backgrounds where a jet is misidentified as the $\tauhadvis$ of the event.  `Other' refers to the
sum of minor backgrounds (vector boson + jets, diboson and Higgs boson); it is primarily composed of
Z$\rightarrow \tau\tau$ in association with light-flavour quarks and $W \to \ell\nu~+~$jets
events. The total background is not identical to the sum of the individual components since the
latter are rounded for presentation, while the sum is calculated with the full precision before
being subsequently rounded. Systematic uncertainties are included. Due to the large correlations,
individual uncertainties can be significantly larger than the total uncertainty. }
\label{tab:yields}
\begin{tabular}{ 
l
S[table-format=4.1]@{\(\;\pm\;\)}S[table-format=2.1]
S[table-format=3.1]@{\(\;\pm\;\)}S[table-format=2.1]
}
\toprule
& \multicolumn{2}{c}{{\lephad\ channel}} & \multicolumn{2}{c}{{\hadhad\ channel}} \\[0.2em]
\toprule
\ttbar     & 2430 & 110 & 94 & 12\\
single-top & 365  & 26 & 20 & 5\\
Fake \tauhad (top) & 140 & 100 & 36 & 11\\
Z$\rightarrow \tau\tau$ + (bb, bc, cc) & 13.1 & 2.7 & 10.1 & 1.3\\
Multi-jet  & \multicolumn{2}{c}{\phantom{i}--} & 30 & 16\\
Other & 91 & 35 & 18 & 7\\
\hline
Total Background  & 3040 & 60 & 207 & 14\\
Data & \multicolumn{2}{l}{3031} & \multicolumn{2}{l}{211}\\

\toprule
\end{tabular}
\end{table}
 
Since no significant excess is observed, upper limits on the scalar and vector LQ pair production
cross-sections for each mass hypothesis are computed based on the modified frequentist
CL$_{\text{s}}$ method~\cite{Read:2002hq}, using a profile likelihood test
statistic~\cite{Cowan:2010js} under the asymptotic approximation.  The resulting observed and
expected limits, assuming $\BR = 1$, as a function of \mlq\ at 95\% confidence level (CL) are shown
in \Fig{\ref{fig:limits1d}} for all LQ models.  The expected contributions of the \lephad\
and \hadhad\ channels are approximately equal at high $\mlq$, while the \hadhad\ is up to a factor
of two more sensitive at low $\mlq$.  The improvement in the observed limit compared with the
expectation is driven by the data deficit in the highest $\hadhad$ PNN score bin mentioned above and
is larger at high $\mlq$ since the signal becomes more localised at high PNN score as $\mlq$
increases. The theoretical prediction for the cross-section of scalar or vector LQ pair production
is indicated by the solid line along with its uncertainties.
 
\begin{table}[tbh]
\begin{center}
\caption{Observed and expected lower limits on the LQ mass at 95\% CL for the three different LQ
models, assuming $\BR = 1$.\protect\label{tab:masslimits}}
\begin{tabular}{lcc}
\toprule
& Obs. limit [\GeV] & Exp. limit [\GeV]\\
\midrule
 
Scalar LQ & 1460 & 1410\\
Vector LQ (minimal-coupling) & 1650 & 1590 \\
Vector LQ (Yang--Mills) & 1910 & 1820 \\
 
\bottomrule
\end{tabular}
\end{center}
\end{table}

The corresponding expected and observed 95\% CL lower limits on the LQ mass for the three different
LQ models are summarised in Table~{\ref{tab:masslimits}}, providing an improvement in mass reach for
a scalar LQ of more than 450\,\GeV\ compared with the previous 36\,\ifb\ result in this
channel~\cite{EXOT-2017-30}.  They extend the full Run~2 ATLAS reach for third-generation up-type
LQs by around 200\,\GeV\ in all three models compared with the search in the $LQLQ \to t\nu t\nu$ decay
mode~\cite{PreviousSUSY}.
 
The results are also expressed as upper limits on the branching ratio to charged leptons as a
function of \mlq\ for each LQ model in \Fig{\ref{fig:limit2D}}.  For all models investigated,
constraints on the LQ mass are reduced by no more than 15\% going from $\BR=1$ to $\BR = 0.5$, while
scalar LQ masses up to around 850\,\GeV\ are excluded for couplings into charged leptons as low as
0.1; the corresponding $\BR=0.1$ exclusion for vector LQ is around 1100\,\GeV\ (1300\,\GeV) in the
minimal-coupling (Yang--Mills) scenario.

\begin{figure}
\centering
\subfigure[]{\includegraphics[width=.49\textwidth]{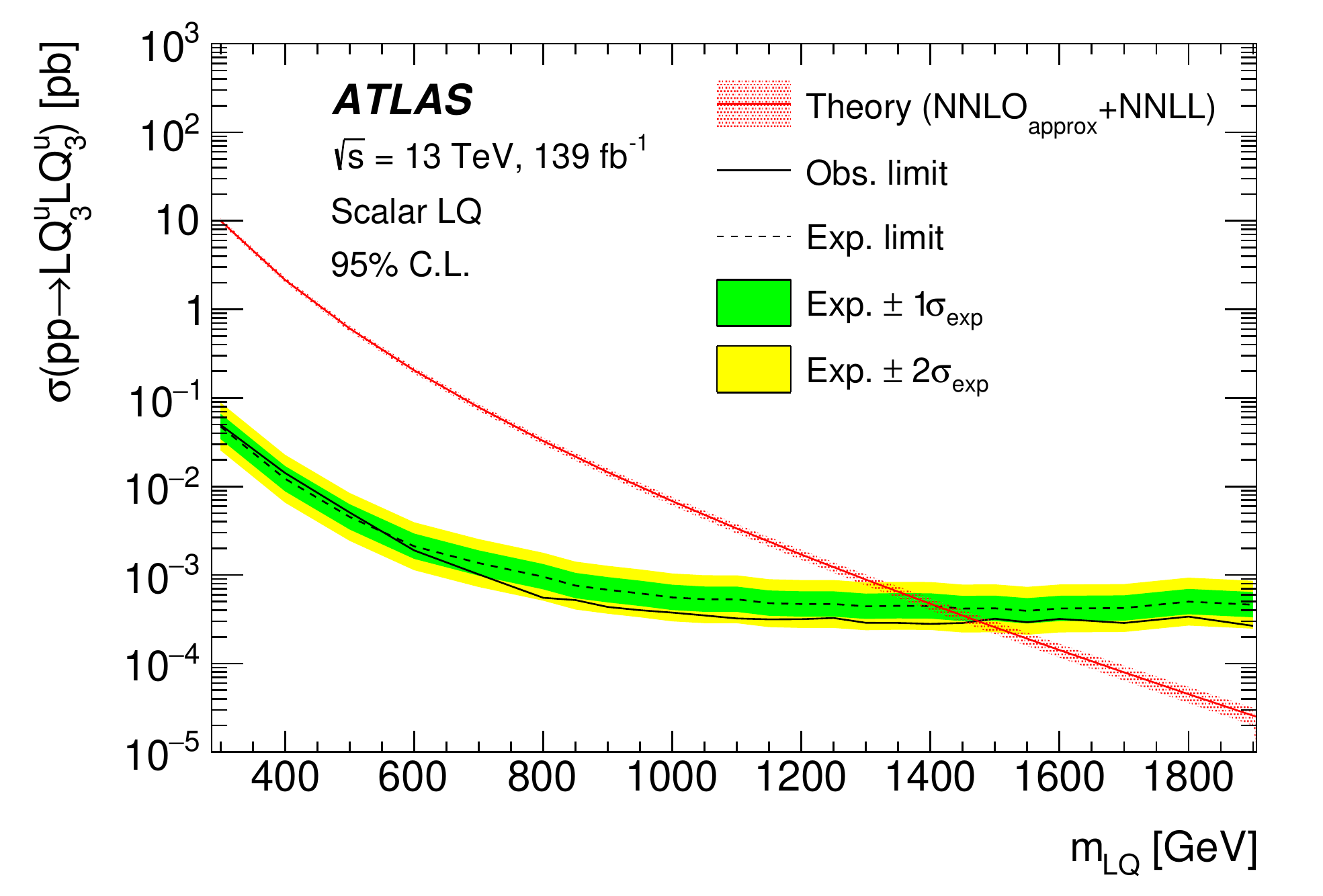}}\\
\subfigure[]{\includegraphics[width=.49\textwidth]{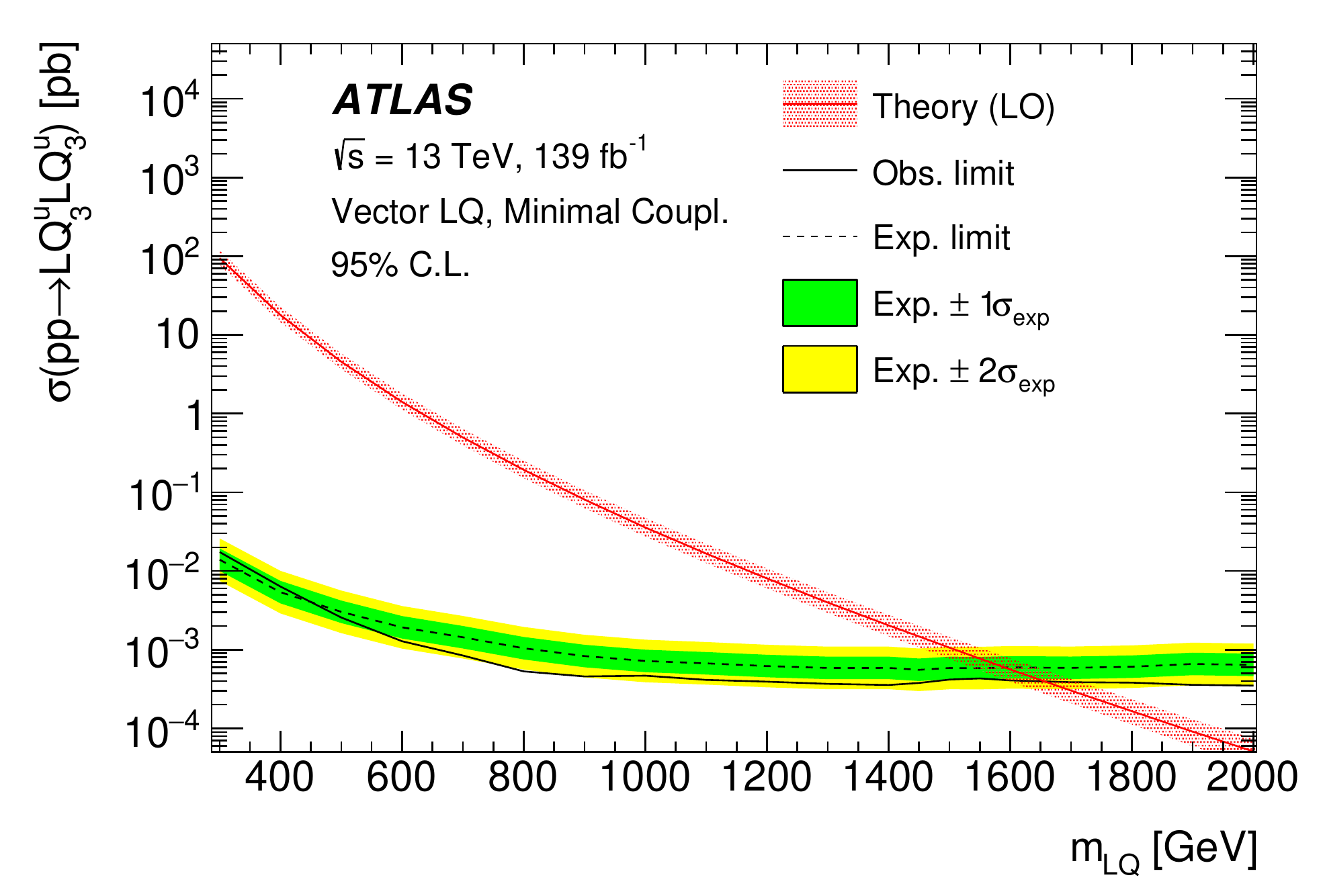}}
\subfigure[]{\includegraphics[width=.49\textwidth]{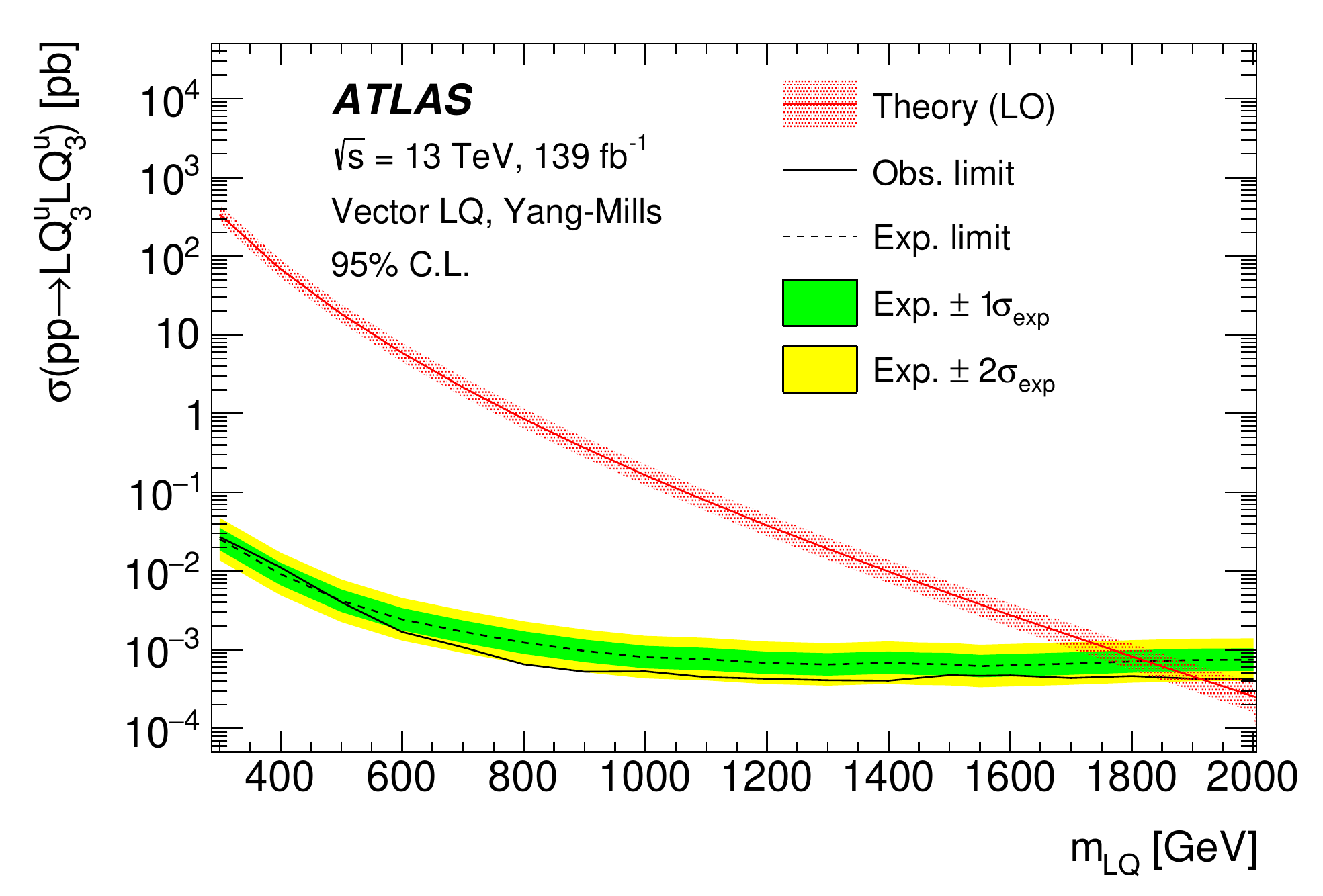}}
\caption{The observed (solid line) and expected (dashed line) 95\% CL upper limits on the LQ pair production cross-sections
assuming $\BR = 1$ as a function of \mlq\ for (a) the scalar LQ case, (b) the vector LQ case in the minimal-coupling
scenario, (c) vector LQs in the Yang--Mills scenario.  The surrounding shaded bands
correspond to the $\pm 1$ and $\pm 2 $ standard deviation ($\pm 1 \sigma, \pm 2 \sigma$)
uncertainty in the expected limit.  The theoretical prediction in each model, along with its
uncertainty, is shown by the lines with the hatched bands. }
\label{fig:limits1d}
\end{figure}
 
\begin{figure}
\centering
\subfigure[]{\includegraphics[width=0.49\textwidth]{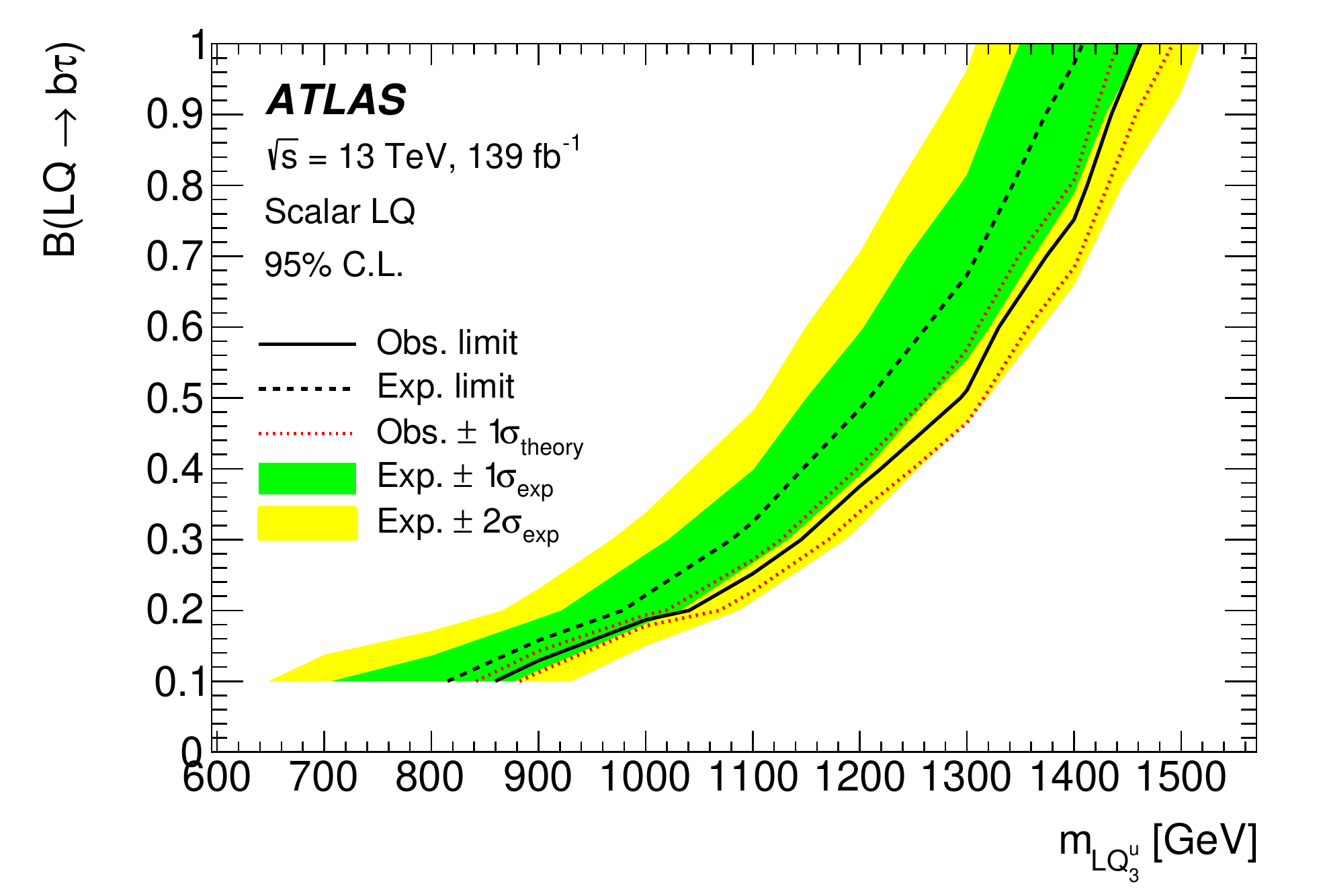}}\\
\subfigure[]{\includegraphics[width=0.49\textwidth]{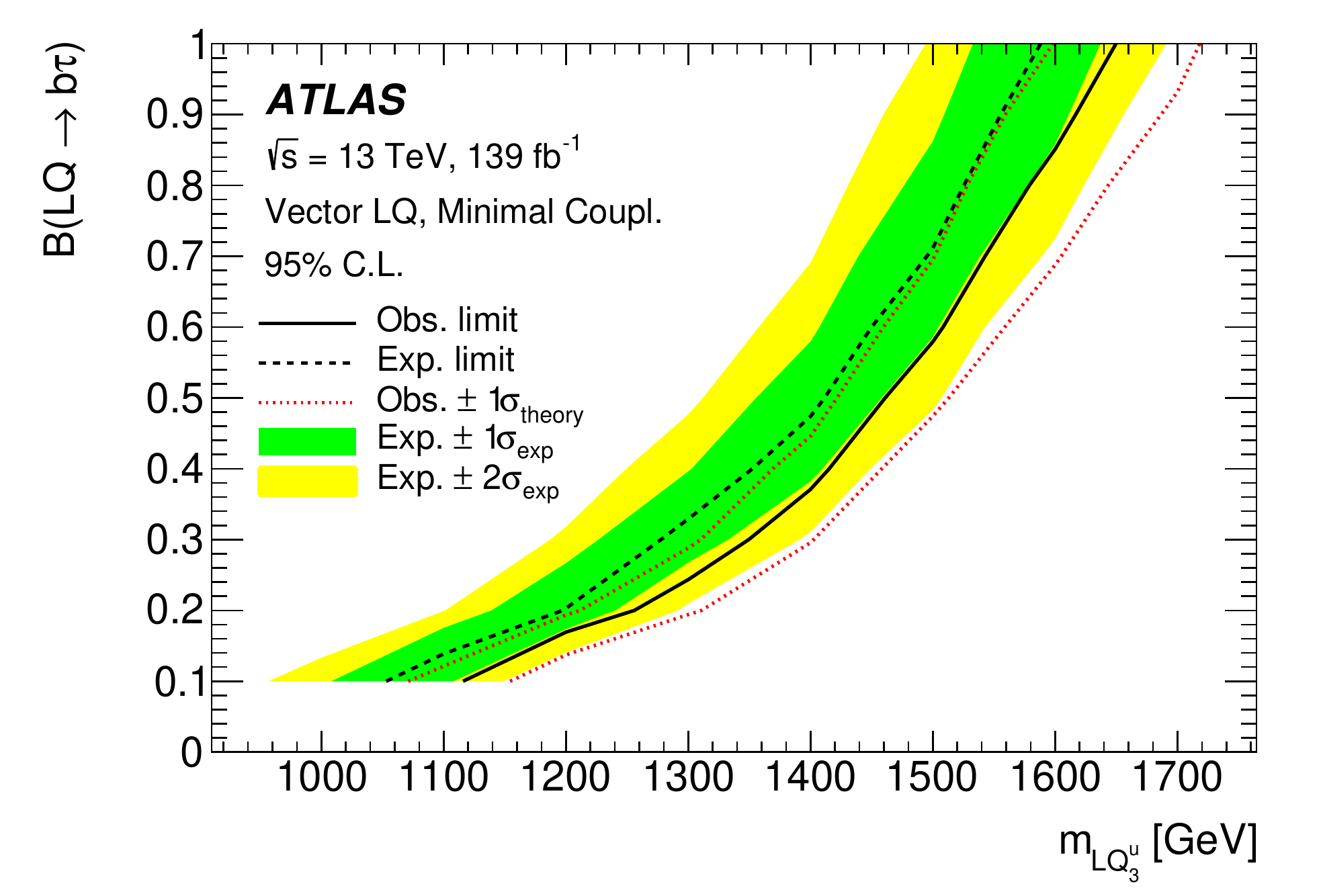}}
\subfigure[]{\includegraphics[width=0.49\textwidth]{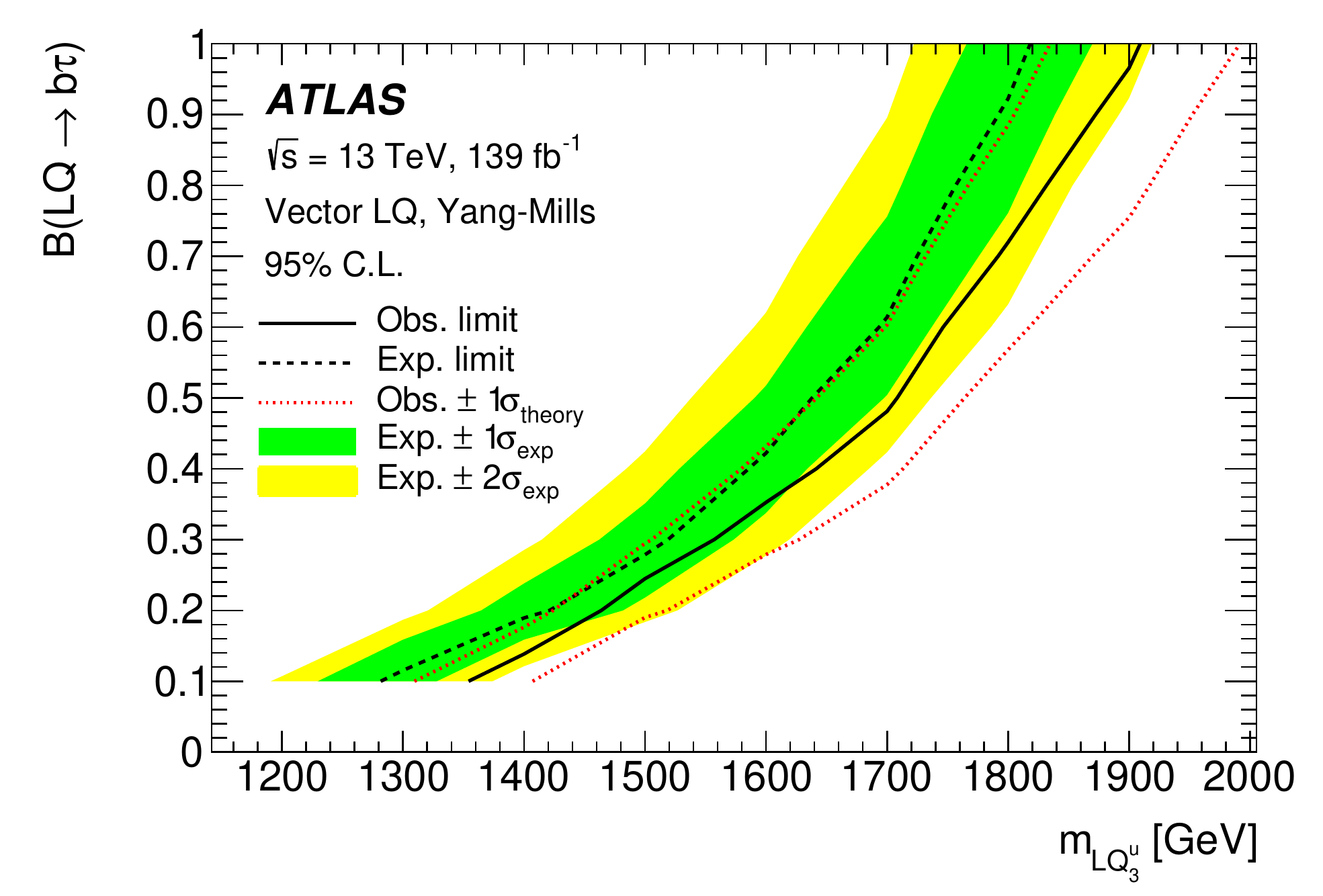}}
\caption{The observed (solid line) and expected (dashed line) 95\% CL upper limits on the branching ratio into charged leptons
as a function of \mlq\ for (a) the scalar LQ case, (b) the vector LQ case in the minimal-coupling scenario, (c)
vector LQs in the Yang--Mills scenario. The observed exclusion region is above the solid line,
with the theoretical uncertainty in the model indicated by the dotted lines around this.  The
expected limit is indicated by the dashed line and the surrounding shaded bands correspond to the $\pm
1$ and $\pm 2 $ standard deviation ($\pm 1 \sigma, \pm 2 \sigma$) uncertainty in the expected
limit. No limits are presented for $\BR < 0.1$ due to the lack of expected signal events in this
final state. }
\label{fig:limit2D}
\end{figure}


\FloatBarrier
\section{Conclusion}
\label{sec:conclusion}

A search for pair-produced scalar or vector leptoquarks decaying into a $b$-quark and a $\tau$-lepton
is presented.  The analysis exploits the full data sample recorded with the ATLAS detector in Run 2 of
the LHC, corresponding to 139\,fb$^{-1}$ of proton–proton collisions at $\sqrt{s} = 13$\,\TeV.  No
significant deviations from the Standard Model expectation are observed and upper limits on the
production cross-section are derived as a function of LQ mass and branching ratio into a
charged lepton.
Scalar LQs with masses below 1460\,\GeV\ are excluded assuming a 100\% branching ratio, while for
vector LQs the corresponding limit is 1650\,\GeV\ (1910\,\GeV) in the minimal-coupling (Yang--Mills)
scenario.  For branching ratios as low as 10\%, scalar LQ masses below around 860\,\GeV\ are
excluded; the corresponding mass limits for vector LQs are 1120\,\GeV\ (1360\,\GeV) in the
minimal-coupling (Yang--Mills) scenario.
These results significantly improve the sensitivity compared to previous ATLAS LQ searches, extending
the mass reach for third-generation up-type LQs by more than 200\,\GeV\ in all models and surpassing
the previous ATLAS search in this final state by more than 450\,\GeV\ for scalar LQs.  In addition to
the increased luminosity, this is due to upgraded hadronic $\tau$-lepton and $b$-jet identification, improved
multivariate techniques and better background estimation methods.


\section*{Acknowledgements}


We thank CERN for the very successful operation of the LHC, as well as the
support staff from our institutions without whom ATLAS could not be
operated efficiently.
 
We acknowledge the support of
ANPCyT, Argentina;
YerPhI, Armenia;
ARC, Australia;
BMWFW and FWF, Austria;
ANAS, Azerbaijan;
CNPq and FAPESP, Brazil;
NSERC, NRC and CFI, Canada;
CERN;
ANID, Chile;
CAS, MOST and NSFC, China;
Minciencias, Colombia;
MEYS CR, Czech Republic;
DNRF and DNSRC, Denmark;
IN2P3-CNRS and CEA-DRF/IRFU, France;
SRNSFG, Georgia;
BMBF, HGF and MPG, Germany;
GSRI, Greece;
RGC and Hong Kong SAR, China;
ISF and Benoziyo Center, Israel;
INFN, Italy;
MEXT and JSPS, Japan;
CNRST, Morocco;
NWO, Netherlands;
RCN, Norway;
MEiN, Poland;
FCT, Portugal;
MNE/IFA, Romania;
MESTD, Serbia;
MSSR, Slovakia;
ARRS and MIZ\v{S}, Slovenia;
DSI/NRF, South Africa;
MICINN, Spain;
SRC and Wallenberg Foundation, Sweden;
SERI, SNSF and Cantons of Bern and Geneva, Switzerland;
MOST, Taiwan;
TENMAK, T\"urkiye;
STFC, United Kingdom;
DOE and NSF, United States of America.
In addition, individual groups and members have received support from
BCKDF, CANARIE, Compute Canada and CRC, Canada;
PRIMUS 21/SCI/017 and UNCE SCI/013, Czech Republic;
COST, ERC, ERDF, Horizon 2020 and Marie Sk{\l}odowska-Curie Actions, European Union;
Investissements d'Avenir Labex, Investissements d'Avenir Idex and ANR, France;
DFG and AvH Foundation, Germany;
Herakleitos, Thales and Aristeia programmes co-financed by EU-ESF and the Greek NSRF, Greece;
BSF-NSF and MINERVA, Israel;
Norwegian Financial Mechanism 2014-2021, Norway;
NCN and NAWA, Poland;
La Caixa Banking Foundation, CERCA Programme Generalitat de Catalunya and PROMETEO and GenT Programmes Generalitat Valenciana, Spain;
G\"{o}ran Gustafssons Stiftelse, Sweden;
The Royal Society and Leverhulme Trust, United Kingdom.
 
The crucial computing support from all WLCG partners is acknowledged gratefully, in particular from CERN, the ATLAS Tier-1 facilities at TRIUMF (Canada), NDGF (Denmark, Norway, Sweden), CC-IN2P3 (France), KIT/GridKA (Germany), INFN-CNAF (Italy), NL-T1 (Netherlands), PIC (Spain), ASGC (Taiwan), RAL (UK) and BNL (USA), the Tier-2 facilities worldwide and large non-WLCG resource providers. Major contributors of computing resources are listed in Ref.~\cite{ATL-SOFT-PUB-2021-003}.


\printbibliography
 
\clearpage
 
\begin{flushleft}
\hypersetup{urlcolor=black}
{\Large The ATLAS Collaboration}

\bigskip

\AtlasOrcid[0000-0002-6665-4934]{G.~Aad}$^\textrm{\scriptsize 102}$,
\AtlasOrcid[0000-0002-5888-2734]{B.~Abbott}$^\textrm{\scriptsize 120}$,
\AtlasOrcid[0000-0002-1002-1652]{K.~Abeling}$^\textrm{\scriptsize 55}$,
\AtlasOrcid[0000-0002-8496-9294]{S.H.~Abidi}$^\textrm{\scriptsize 29}$,
\AtlasOrcid[0000-0002-9987-2292]{A.~Aboulhorma}$^\textrm{\scriptsize 35e}$,
\AtlasOrcid[0000-0001-5329-6640]{H.~Abramowicz}$^\textrm{\scriptsize 151}$,
\AtlasOrcid[0000-0002-1599-2896]{H.~Abreu}$^\textrm{\scriptsize 150}$,
\AtlasOrcid[0000-0003-0403-3697]{Y.~Abulaiti}$^\textrm{\scriptsize 117}$,
\AtlasOrcid[0000-0003-0762-7204]{A.C.~Abusleme~Hoffman}$^\textrm{\scriptsize 137a}$,
\AtlasOrcid[0000-0002-8588-9157]{B.S.~Acharya}$^\textrm{\scriptsize 69a,69b,p}$
% ...
\bigskip
\\

% ...
$^{29}$Physics Department, Brookhaven National Laboratory, Upton NY; United States of America.\\
% ...
$^{35}$$^{(a)}$Facult\'e des Sciences Ain Chock, R\'eseau Universitaire de Physique des Hautes Energies - Universit\'e Hassan II, Casablanca;$^{(b)}$Facult\'{e} des Sciences, Universit\'{e} Ibn-Tofail, K\'{e}nitra;$^{(c)}$Facult\'e des Sciences Semlalia, Universit\'e Cadi Ayyad, LPHEA-Marrakech;$^{(d)}$LPMR, Facult\'e des Sciences, Universit\'e Mohamed Premier, Oujda;$^{(e)}$Facult\'e des sciences, Universit\'e Mohammed V, Rabat;$^{(f)}$Institute of Applied Physics, Mohammed VI Polytechnic University, Ben Guerir; Morocco.\\
% ...
$^{55}$II. Physikalisches Institut, Georg-August-Universit\"{a}t G\"ottingen, G\"ottingen; Germany.\\
% ...
$^{69}$$^{(a)}$INFN Gruppo Collegato di Udine, Sezione di Trieste, Udine;$^{(b)}$ICTP, Trieste;$^{(c)}$Dipartimento Politecnico di Ingegneria e Architettura, Universit\`a di Udine, Udine; Italy.\\
% ...
$^{102}$CPPM, Aix-Marseille Universit\'e, CNRS/IN2P3, Marseille; France.\\
% ...
$^{117}$Department of Physics, New York University, New York NY; United States of America.\\
% ...
$^{120}$Homer L. Dodge Department of Physics and Astronomy, University of Oklahoma, Norman OK; United States of America.\\
% ...
$^{137}$$^{(a)}$Departamento de F\'isica, Pontificia Universidad Cat\'olica de Chile, Santiago;$^{(b)}$Millennium Institute for Subatomic physics at high energy frontier (SAPHIR), Santiago;$^{(c)}$Instituto de Investigaci\'on Multidisciplinario en Ciencia y Tecnolog\'ia, y Departamento de F\'isica, Universidad de La Serena;$^{(d)}$Universidad Andres Bello, Department of Physics, Santiago;$^{(e)}$Instituto de Alta Investigaci\'on, Universidad de Tarapac\'a, Arica;$^{(f)}$Departamento de F\'isica, Universidad T\'ecnica Federico Santa Mar\'ia, Valpara\'iso; Chile.\\
% ...
$^{150}$Department of Physics, Technion, Israel Institute of Technology, Haifa; Israel.\\
$^{151}$Raymond and Beverly Sackler School of Physics and Astronomy, Tel Aviv University, Tel Aviv; Israel.\\
% ...
$^{a}$ Also Affiliated with an institute covered by a cooperation agreement with CERN.\\
$^{b}$ Also at An-Najah National University, Nablus; Palestine.\\
$^{c}$ Also at Borough of Manhattan Community College, City University of New York, New York NY; United States of America.\\
$^{d}$ Also at Bruno Kessler Foundation, Trento; Italy.\\
$^{e}$ Also at Center for High Energy Physics, Peking University; China.\\
$^{f}$ Also at Center for Interdisciplinary Research and Innovation (CIRI-AUTH), Thessaloniki; Greece.\\
% ...
$^{p}$ Also at Department of Physics, King's College London, London; United Kingdom.\\
% ...

\end{flushleft}


\end{document}